\documentclass[letterpaper, 11pt]{article}

\usepackage[nocompress]{cite}
\usepackage{jheppub}

\usepackage{graphicx}
\usepackage{epstopdf}
\usepackage{amsmath, amssymb}
\usepackage{bbm}

\usepackage{caption}
\usepackage{subcaption}

\usepackage{multirow} 
\usepackage{longtable} 

\usepackage{romannum}

\usepackage{harpoon}
\usepackage[vcentermath]{youngtab}

\usepackage{tikz, tikz-3dplot, pgfplots}
\usepackage{tkz-graph}
\usetikzlibrary[positioning,patterns] 

\usepackage[us,12hr]{datetime} 

\newcommand{\be}{\begin{eqnarray}}
\newcommand{\ee}{\end{eqnarray}}

\newcommand{\bn}{\begin{enumerate}}
\newcommand{\en}{\end{enumerate}}

\def\CD{{\cal D}}

\def\CL{{\cal L}}

\def\CN{{\cal N}}
\def\CO{{\cal O}}

\def\CR{{\cal R}}
\def\CS{{\cal S}}
\def\CT{{\cal T}}

\def\Tr{{\rm Tr}}
\def\tr{{\rm Tr}}

\newcommand{\bea}{\begin{eqnarray}}
\newcommand{\eea}{\end{eqnarray}}

\allowdisplaybreaks

\newcommand{\ba}[1]{\begin{align} #1 \end{align} }

\newcommand{\bs}[1]{\begin{split} #1 \end{split} }

\def\tr{{\text{Tr}}}

\title{Dualities of Adjoint SQCD and Supersymmetry Enhancement}

\affiliation[a]{Faculty of Science and Technology, Seikei University\\
3-3-1 Kichijoji-Kitamachi, Musashino-shi, Tokyo, 180-8633, Japan}
\affiliation[b]{Nambu Yoichiro Institute of Theoretical and Experimental Physics, Osaka Metropolitan University\\3-3-138, Sugimoto, Sumiyoshi-ku, Osaka, 558-8585, Japan}
\affiliation[c]{Kavli Institute for the Physics and Mathematics of the Universe\\ University of Tokyo, Kashiwa, Chiba 277-8583, Japan}
\affiliation[d]{Department of Physics, Korea Advanced Institute of Science and Technology\\
291 Daehak-ro, Yuseong-gu, Daejeon 34141, Korea}

\author[a,b]{Kazunobu Maruyoshi,}
\author[c]{Emily Nardoni,}
\author[d]{and Jaewon Song}

\emailAdd{maruyoshi@st.seikei.ac.jp}
\emailAdd{emily.nardoni@ipmu.jp}
\emailAdd{jaewon.song@kaist.ac.kr}

\abstract
{
We propose a new dual description of four-dimensional $\CN=1$ $SU(N)$ gauge theory with one adjoint ($X$) and $N_f$ fundamental matters with a superpotential $W = \Tr X^{p+1}$. The dual theory consists of the $\CD_p[SU(N)]$ Argyres-Douglas theory coupled to $SU(N)$ gauge theory and $N_f$ fundamentals with a superpotential deformation. We study renormalization group fixed points of the Argyres-Douglas dual theories with and without superpotential deformations, and identify the conditions for them to be dual to the fixed points of adjoint SQCD. We check our proposal via matching central charges, chiral operators and superconformal indices. We find that when $N_f = 2N$ and $p=2$, the dual theory flows to $\CN=2$ $SU(N)$ superconformal QCD with $2N$ flavors upon suitable superpotential deformation, exhibiting supersymmetry enhancement. 
}

\begin{document}
\maketitle

\def\arraystretch{1}

\section{Introduction}

One of the many achievements of supersymmetry has been to shine a light on a host of nontrivial, interacting superconformal fixed points in four and greater dimensions. In many interesting cases, these superconformal field theories (SCFTs) exhibit dualities, which then provide a powerful window into their strong coupling dynamics. In the present context, a duality between two field theories means that they belong to the same universality class, flowing at low energies to the same fixed point, and thus describing precisely the same IR physics.  Duality is an especially useful tool when it relates a strongly-coupled quantum field theory (QFT) to a weakly-coupled one; in such cases, the weakly-coupled dual can be fruitfully used to ascertain properties of the QFT that were inaccessible from the original description. The quintessential example of such a supersymmetric duality is Seiberg duality of $\CN=1$ supersymmetric QCD in four dimensions \cite{Seiberg:1994pq}, which illuminates the IR phase of the theory in the region that the magnetic dual is weakly coupled.

In this work we revisit $\CN=1$ adjoint supersymmetric QCD in four dimensions, with gauge group $SU(N)$, $N_f$ flavors of fundamental and antifundamental chiral multiplets, and one adjoint chiral multiplet $X$. The interacting SCFTs that can be reached via renormalization group (RG) flow from this class of gauge theories are classified by their superpotential \cite{Kutasov:1995ve,Kutasov:1995np,Kutasov:1995ss},
\ba{
W_{A_k} = \tr \, X^{k+1}\,,\label{wa}
}
labeled for the identification of the polynomial in $X$ with the $A_k$-type singularity. (More generally, when allowing for the maximum number of  two adjoint chiral multiplets compatible with asymptotic freedom, the possible SCFTs are labeled by Arnold's ADE simple surface singularities \cite{Intriligator:2003mi}.)
The $W_{A_k}$ SCFTs possess conjectured Seiberg-like dualities (called as Kutasov-Schwimmer dualities) within a conformal window of $N$ and $N_f$ \cite{Kutasov:1995ve,Kutasov:1995np}, where the dual has $SU(k N_f-N)$ gauge group and additional superpotential terms. Much as in Seiberg's original duality, fundamental fields in one description map to composite operators in the other, and strong and weak coupling are interchanged.

We propose a new dual description of the $W_{A_k}$ SCFTs, consisting of a strongly-coupled, inherently non-Lagrangian $\CN=2$ SCFT coupled to $\CN=1$ vector and chiral multiplets. The non-Lagrangian sector of this dual is of Argyres-Douglas type, named as such for the seminal example of an interacting SCFT with relatively non-local degrees of freedom, originally found by tuning to special points on the moduli space of $\CN=2$ gauge theories \cite{Argyres:1995jj}. The Argyres-Douglas theories that play a role in our constructions are the $\CN=2$ $\CD_p[G]$ SCFTs \cite{Xie:2012hs,Cecotti:2012jx,Cecotti:2013lda}, which can be constructed by compactifying the 6d (2,0) SCFTs on a sphere with an irregular puncture \cite{Bonelli:2011aa, Xie:2012hs, Wang:2015mra}. 
Key to the proposed duality is the observation that the $\CD_p[G]$ theories behave (in a sense that we will make precise) like chiral multiplets that transform in the adjoint representation of $G$, and have fractional scaling dimension upon a suitable superpotential deformation \cite{Bolognesi:2015wta, Xie:2021omd, Kang:2023pot, Bajeot:2023gyl}. There have been multiple clues that the $\CD_p[G]$ theories behave like free hypermultiplets (especially for $p=2$), as observed {\it e.g.} via the similarities of their Schur indices \cite{Xie:2016evu, Song:2017oew, Buican:2017fiq, Buican:2017rya}, and our duality can be thought of as another incarnation of the ``simplicity" of the $\CD_p[G]$ theories. More precisely, we find that upon $\CN=1$ gauging and deformation, the $\CD_{p=k}[G]$ theories behave like chiral multiplets in the adjoint of $G$ with a superpotential \eqref{wa} of ${A_k}$ type.

Below we perform a variety of checks of the duality, including the matching of the central charges and 't Hooft anomalies, global symmetries, and superconformal indices in the Schur limit. 
The mapping between the two sides of the duality is nontrivial; for instance, the truncated mapping between the mesons and baryons of adjoint SQCD and the proposed dual follows from a nontrivial Higgs branch relation in the Argyres-Douglas SCFT. The duality is proposed to hold within a certain conformal window of $N$ and $N_f$ which depends on the order $k$ of the superpotential, and which we carefully determine below.  In the special case of the $SU(3)$ theory with cubic $\tr \, X^3$ superpotential, we demonstrate that the full superconformal indices on both sides of the duality match.

On the one hand, the proposed duality is a new example of a Lagrangian dual for a non-Lagrangian QFT (in the spirit of  {\it e.g.} \cite{Gadde:2015xta, Maruyoshi:2016tqk, Maruyoshi:2016aim,  Agarwal:2016pjo, Agarwal:2017roi, Benvenuti:2017bpg,  Agarwal:2018ejn, Razamat:2019vfd, Zafrir:2019hps, Razamat:2020gcc, Zafrir:2020epd, Etxebarria:2021lmq, Kang:2023pot, Bajeot:2023gyl}), providing the opportunity to glean new insights into the strongly-coupled dual. For example, it suggests a quantum truncation of the chiral ring of the $\CN=1$ deformed Argyres-Douglas theory, much as the proposed Seiberg-like dualities for the $D$ and $E$-series generalizations of the $W_{A_k}$ SCFTs rely on a proposed quantum truncation \cite{Brodie:1996vx, Kutasov:2014yqa} (see also \cite{Intriligator:2016sgx}). On the other, our proposal suggests the possibility for a zoo of new dualities of strongly-coupled $\CN=1$ gauge theories built from deformed $\CD_p[G]$ theories, which will be interesting to explore. 

We highlight that when $p=2$ with $N_f=2N$, our dual theory (further deformed by a superpotential term) flows to superconformal QCD with $\CN=2$ supersymmetry, thereby exhibiting supersymmetry enhancement. The supersymmetry enhancement works in a very analogous fashion as to the cases studied in \cite{Kang:2023pot, Bajeot:2023gyl}, except that we need a superpotential term to trigger the RG flow to the $\CN=2$ fixed point. This observation can be easily generalized to quiver gauge theories, and would be interesting to explore further in this context.

\paragraph{Outline}

The outline for the rest of this paper is as follows. In section \ref{sec:Ak}, we discuss $\CN=1$ adjoint SQCD in four dimensions and its deformation by the superpotential \eqref{wa}, reviewing the properties of the $W_{A_k}$ SCFTs. In section \ref{sec:dual}, we describe the $\CN=1$ theory which is proposed to be dual to  adjoint SQCD. After studying the properties of the fixed point of the dual theory, we perform  various checks of the duality, including the matching of gauge-invariant operators, central charges, and superconformal indices. We conclude in section \ref{sec:conclusion} with a discussion on the possible generalization of the duality to $D$-type and $E$-type SCFTs. In appendix \ref{sec:scft}, we fix our notation for the $\CN=1$ and $\CN=2$ supersymmetries. Appendix \ref{sec:Dp} is devoted to a review of the $\CD_p[SU(N)]$ Argyres-Douglas theories.

\section{The \texorpdfstring{$W_{A_k}$}{WAk} fixed points of \texorpdfstring{$\mathcal{N}=1$}{N=1} adjoint SQCD}
\label{sec:Ak}

In this section we review the salient features of $\CN=1$ $SU(N)$ adjoint SQCD in four dimensions, deformed by the $W_{A_k}$ superpotential \eqref{wa}. 

\subsection{Preliminaries }

We consider the four-dimensional $SU(N)$ gauge theory consisting of an $\CN=1$ vector multiplet, $N_f$ chiral multiplets $Q$ and $\widetilde{Q}$ transforming in the fundamental and antifundamental representations of the gauge group (the quarks), and one chiral multiplet $X$ transforming in the adjoint representation of the gauge group. We refer to the IR fixed point of the theory without superpotential as the $\widehat{A}$ SCFT, following the notation of \cite{Intriligator:2003mi}.\footnote{~This gauge theory generically flow to a product CFT where some of the gauge-invariant operators decouple along the RG flow \cite{Kutasov:2003iy, Agarwal:2019crm, Agarwal:2020pol}.} 

We deform by the following superpotential for the adjoint $X$,
\ba{
W_{A_k} = \tr \, X^{k+1}\,.\label{wa2}
}
In the case of $k=1$, the superpotential gives a mass to $X$, and at low energies the resulting theory is SQCD (with no adjoint). For $k>1$, \eqref{wa2} is relevant at the $\widehat{A}$ fixed point only if the number of flavor $N_f$ is below a maximum value that depends on $k$, which we review in some detail below. In this range, the superpotential drives the theory to a new fixed point that we will refer to as the $W_{A_k}$ SCFT. The phase structure of the theory for smaller values of $N_f$ is elucidated by the magnetic dual,\footnote{~Although, note that there is no known dual for the $\widehat{A}$ fixed point, and we will not suggest one here.} an $SU(kN_f-N)$ gauge theory which becomes more weakly coupled as the original electric description becomes more strongly coupled \cite{Kutasov:1995ve,Kutasov:1995np,Kutasov:1995ss}.

\def\arraystretch{1.3}
\begin{table}[t!]
\centering
\begin{tabular}{|c||c|c|c|c|c|}
\hline
 & $SU(N_f)_L$ & $SU(N_f)_R$ & $U(1)_{B}$  & $U(1)_{R}$ at $W_{A_k}$    \\ \hline \hline
$Q$  & $\Box$ &1    & $1/N$   & $1-\frac{N}{N_f} \frac{2}{k+1}$     \\ \hline
$\widetilde{Q}$  & 1 & $\overline{\Box}$   & $-1/N$     & $1-\frac{N}{N_f}\frac{2}{k+1}$     \\ \hline
$X$  &1  & 1 &     0 & $\frac{2}{k+1}$    \\ \hline
$W_{\alpha}$  & 1 & 1 & 0 &       1      \\ \hline \hline
$\tr \, X^{j}$ & 1 & 1 & 0 & $\frac{2j}{k+1}$ \\ \hline
$M_j$ & $\Box$ & $\overline{\Box}$ & 0 & $2\left(1 +  \frac{ j - 1-2N/N_f}{k+1}\right) $  \\ \hline
$B^{(n_1,\dots,n_k)}$ & 
\def\arraystretch{0.8}
\footnotesize
$\begin{array}{c} N\text{-index} \\  \text{antisym}  \end{array}$ \def\arraystretch{1} & 1 &  1 & $\frac{2}{k+1}\left( \sum_{\ell=1}^k n_\ell \ell + N \left( \frac{k-1}{2} - \frac{N}{N_f} \right)   \right)$ \\ \hline
\hline\end{tabular}
\caption{The charges of the fields under the global symmetries, with $U(1)_R$ given at the $W_{A_k}$ fixed point. The lower part of the table lists the charges of the gauge-invariant operators listed in \eqref{giop}.  
\label{tab:Akecharges1}
}
\end{table}
\def\arraystretch{1}

The anomaly-free global continuous symmetry is 
\ba{
SU(N_f)_L\times SU(N_f)_R \times U(1)_B \times U(1)_R\,,
}
under which the microscopic fields have charges given in Table \ref{tab:Akecharges1}.\footnote{~There are finite symmetries which will not play a role in our discussion. The theory without superpotential has an additional non-anomalous $U(1)$ under which $X,Q,\widetilde{Q}$ transform, which is broken by \eqref{wa2}.}  The chiral ring of operators consists of the gauge-invariant products of these fields, subject to relations from {\it e.g.} the superpotential.
A basis of adjoint-valued products is given by operators $X^{j-1}$, $j=1,\dots,\alpha$. For the $\widehat{A}$ theory,  $\alpha=N$ due to a classical matrix relation for $X$ that relates the operator $X^N$ to lower powers $X^{\ell < N}$,\footnote{~For instance, for $SU(2)$ one has $X^2 = \tfrac{1}{2}\mathbbm{1}\,\tr X^2$, while for $SU(3)$, $X^3 = \tfrac{1}{2} X\, \tr X^2 + \tfrac{1}{3}\mathbbm{1}\, \tr X^3$. 
} while the $W_{A_k}$ case is classically truncated at $\alpha=k$ due to the superpotential $F$-term. Gauge-invariant operators are given by taking traces, or contracting with the $Q$ and $\widetilde{Q}$ to form dressed quark operators, leading to the following,
\ba{
\label{giop}
\bs{
&\tr \, X^{j}\,,\qquad \qquad\ \   j=2,\dots,\alpha \\
M_j &= \widetilde{Q} X^{j-1} Q\,,\qquad j=1,\dots,\alpha \\
B^{(n_1,\dots,n_j)} &= Q^{n_1}\left( X Q \right)^{n_2}\cdots ( X^{j-1} Q )^{n_j}\,,\quad \sum_{\ell=1}^j n_\ell = N\,; \quad n_\ell \leq N_f\,;\quad j= 1,\dots,\alpha  \,.
}}
We will refer to the $M_j$ as generalized mesons, and the $B^{(n_1,\dots,n_j)}$ as generalized baryons.
The charges of these operators are given in the bottom section of Table \ref{tab:Akecharges1}. 
For a given $j$, there are $(\footnotesize{\begin{array}{c}  j N_f \\ N\end{array}})$ baryons of the form \eqref{giop}, where the $N$ color indices are contracted with an $\epsilon$ tensor, as well as analogous anti-baryons $\widetilde{B}$ composed of the antifundamental $\widetilde{Q}$ fields.

\subsection{Flow from \texorpdfstring{$\widehat{A}\to W_{A_{k}}$ and $a$-maximization review}{Ahat to Ak}}

\label{sec:amax}

The flow from the $\widehat{A}$ fixed point to the $W_{A_k}$ fixed point of adjoint SQCD occurs within a range of $N_f$ that depends on $N$ and $k$, and was analyzed in detail in \cite{Kutasov:2003iy}. Here we review this analysis, with the two-fold motivation of first clarifying the range of parameters for which we expect a conformal window with our proposed Argyres-Douglas dual, and second providing a case study in the procedure of $a$-maximization that will be useful in later sections.

The $a$ and $c$ central charges in a superconformal field theory are given in terms of the superconformal $U(1)_R$ symmetry as \cite{Anselmi_1998},
\ba{
a = \frac{3}{32}\left(3 \tr\, R^3 - \tr \,R \right)\,,\qquad c = \frac{1}{32}\left(9 \tr\, R^3 - 5 \tr \,R \right)\,.
}
The exact superconformal $R$-symmetry at the fixed point can be a mixture of the original $U(1)_R$ with $U(1)$ flavor symmetries. Then, the  R-charges of the fields at a given fixed point are determined by locally maximizing the $a$ central charge over all possible $U(1)_R$ symmetries, in a process known as $a$-maximization \cite{Intriligator:2003jj}. Since the dimensions of chiral primary operators are given in terms of their R-charges as $\Delta_{\mathcal{N}=1}(\CO) = \frac{3}{2}R(\CO)$, this procedure equivalently fixes their scaling dimensions at the fixed point.

Let us see how this works at the $\widehat{A}$ fixed point. The trial $a$-function, given in terms of the R-charges $R_i$ of the scalars in representations $r_i$ of the gauge group $G$, is 
\ba{
\label{trial}
a = \frac{3}{32}\left( 2 |G| + \sum_i |r_i| \left( 3 (R_i - 1)^3 - (R_i-1) \right) \right)\,. 
}
The R-charges of the fields are also constrained by the condition of 
chiral anomaly cancellation $\tr\, U(1)_R SU(N)^2 = 0$, which is equivalent to the vanishing of the $\beta$-function, and implies that the quark and adjoint R-charges are related as,
\ba{
\quad R(Q) = R(\widetilde{Q}) = 1 - \frac{N}{N_f} R(X)\,.
\label{ahatrel}
}
The trial $a$-function \eqref{trial} for the ${\widehat{A}}$ theory evaluates to
\ba{
\bs{
a_{\widehat{A}}= \frac{3}{32}\bigg( 2 (N^2-1) + 2 NN_f \left( 3 (R(Q) - 1)^3 - (R(Q)-1)\right) \\
+ (N^2-1) \left( 3 (R(X) - 1)^3 - (R(X)-1)\right) 
\bigg)\,.
\label{atestahat}
}}
Substituting \eqref{ahatrel} into \eqref{atestahat} and maximizing with respect to $R(X)$ yields
\ba{
R(X) = \frac{  N^2-1 -\tfrac{1}{3} \sqrt{ 1 - N^4 + 20\, {N^6}/{N_f^2}     - 16\, N^4/N_f^2 )}}{N^2 -1- 2\, N^4/N_f^2}\,,
\label{rx}
}
with $R(Q)$ then determined by \eqref{ahatrel}.

These trial R-charges are corrected when a gauge-invariant operator $\mathcal{O}$ hits or falls below the unitarity bound $R(\CO) = 2/3$, signaling that the operator has become free and should be decoupled. Then, $a$ is corrected by
\ba{
\label{acorr}
a_{\text{corr}} =- \frac{3}{32} \left( 3 (R(\CO) - 1)^3 - (R(\CO)-1) \right)\,.
}
Adding \eqref{acorr} to \eqref{atestahat} and re-maximizing will yield the corrected R-charges. More generally, any accidental $U(1)$ symmetry which emerges at low energies might mix with $U(1)_R$ at the fixed point, and the results of $a$-maximization are only reliable when all such accidental symmetries have been taken into account.

Once all apparent operator decouplings have been taken into account, and assuming no unforeseen accidental symmetries, the results of $a$-maximization at the $\widehat{A}$ fixed point can then be applied to determine whether the deformation by $\tr X^{k+1}$ is relevant and drives the theory to the $W_{A_k}$ fixed point. If this is the case, then the R-charges at the new fixed point are fixed by the superpotential \eqref{wa2} (which must have R-charge 2), and the anomaly-free condition \eqref{ahatrel}, which together enforce that 
\ba{
\label{akrcharges}
R(X) = \frac{2}{k+1}\,,\qquad R(Q) = R(\widetilde{Q}) = 1 - \frac{N}{N_f}\frac{2}{k+1}\,. 
}
The central charges are then determined to be,
\ba{
\label{acak}
\bs{
a_{A_k} &= \frac{3(-12 N^4 + N^2 N_f^2 (5k^2 + k + 2) - N_f^2 (4k^2 - k +1)}{8N_f^2 (k+1)^3} \,,\\
c_{A_k} &= \frac{-36 N^4 + N^2 N_f^2 (16k^2 + 5k + 7) - N_f^2 ( 11k^2 - 5k +2)}{8N_f^2 (k+1)^3} \,.
}}
The R-charges of the microscopic fields and gauge-invariant operators at the $W_{A_k}$ fixed point are given in Table \ref{tab:Akecharges1}. 
We remark that these values of the central charges in \eqref{acak} are generically \emph{not} the true values of central charges at the infrared fixed point. The reason is that generically our gauge theory flows to a product CFT (interacting part times free fields) which results in accidental symmetry, modifying the central charges.

The range of $N_f$ as a function of $N$ for which the flow to the $W_{A_k}$ fixed point occurs in adjoint SQCD is restricted by the following considerations:
\begin{itemize}
\item $N_f \leq 2 N$: $N_f$ is bounded from above by the condition of asymptotic freedom of the gauge theory. 
\item $N_f < (N_f^{\text{max}})_k$: The upper bound on $N_f$ is at most $2N$ due to the previous bullet point, but in general is further restricted by the requirement that $\tr X^{k+1}$ is a relevant operator at the $\widehat{A}$ fixed point. This bound depends on $k$.
In the Veneziano limit of large $N$ and large $N_f$ with fixed ratio $N/N_f$, it was found in 
\cite{Kutasov:2003iy} to obey
\ba{
(N_f^{\text{max}})_k = \frac{6 N}{\sqrt{5k^2 - 8k+5}}\,,\qquad k = 2,\dots,15\,,\qquad N,N_f\to \infty\,,
}
for the first few values of $k$.\footnote{~In this limit, the mesons are the first gauge invariant operators that violate the unitarity bound and whose decoupling affects the large-$N$ R-charges, which must be taken into account for $k>15$.} 
For finite values of $N$ and $N_f$, 
when no operators decouple this bound is determined by checking when $R(X^{k+1}) < 2$ using \eqref{rx}.

\item $N_f > N/k$: $N_f$ is bounded from below by the requirement of a stable vacuum \cite{Kutasov:1995np}. 
\item $N_f > N_f^{\text{decouple}}$: There will also be a lower bound on $N_f$ below which operators decouple along the flow to the $\widehat{A}$ fixed point, which can be more restrictive than the previous bullet point. This decoupling bound follows from checking (using \eqref{rx}) when the smallest dimension operator $\tr X^2$ crosses the unitarity  bound, leading to
\ba{
\label{nfdec}
N_f^{\text{decouple}} = \frac{2 N^2}{\sqrt{10 N^2 - 6}}\,.
}
For values of $N_f$ smaller than $N_f^{\text{decouple}}$, the R-charges must be corrected due to decoupling of the free operators {\it before} 
determining whether  $\tr X^{k+1}$ is a relevant deformation. 
\end{itemize}

To illustrate these bounds, 
in Table \ref{tab:sqcdtable} we give the allowed values of $N/k < N_f < (N_f^{\text{max}})_k$ for which $\tr X^{k+1}$ is relevant and drives a flow to a new fixed point for small values of $N$ and $k$, including (1) 
 which operators decouple at the $\widehat{A}$ fixed point if  $N_f < N_f^{\text{decouple}}$, and (2)  which operators decouple at the $W_{{A}_k}$ fixed point for these values of $N$ and $N_f$.\footnote{~No Casimir operators $\tr X^j$ decouple at the $W_{A_k}$ fixed points for the examples listed in Table \ref{tab:sqcdtable}, but note that for larger values $k\geq 5$ some of these operators will decouple, since $R(X^j) = 2j / (k+1)$.}~\footnote{~Not all of the $W_{A_k}$ fixed points listed in Table \ref{tab:sqcdtable} are interacting (as is obvious from the perspective of the $SU(kN_f-N)$ Kutasov-Schwimmer dual); in the cases of $(N,k)=(3,2)$ with $N_f=2$, $(N,k)=(5,2)$ with $N_f=3$, and  $(N,k)=(5,3)$ with $N_f=2$, the IR theory after deformation by $W_{A_k}$ is free.  
}
We will use these considerations in the next section to examine the conformal window for which the proposed Argyres-Douglas duals describe a flow to the same fixed point.

\def\arraystretch{1.2}
\begin{table}[t!]
\centering
\begin{tabular}{|c||c|c|c||c|c|c||c|c|c||} 
\hline
$(N,k)$ & \multicolumn{3}{c||}{$(3,2)$} & \multicolumn{3}{c||}{$(4,3)$} & \multicolumn{3}{c||}{$(5,2)$}   \\ \hline
$N_f$ & 2 &  3 &  4-6 & 2 &  3 &  4 & 3  & 4-5 & 6-10  \\ \hline 
decouple at $\widehat{A}$ & 
 &  &  &
$\tr X^2$ &   & $\phantom{M_1}$  & 
$\tr X^2$ & &   \\
decouple at ${A}_k$& 
$ M_{1}, M_2, B^{(2,1)}$  & $M_{1}$ &  & 
$M_{1}, M_2$ & $M_{1}$ &   & 
$M_1,M_2$  & $M_1$ & \\ \hline
\end{tabular}

\bigskip

\begin{tabular}{|c||c|c|c||c|c|c||} 
\hline
$(N,k)$ & \multicolumn{3}{c||}{$(5,3)$} & \multicolumn{3}{c||}{$(5,4)$}  \\ \hline
$N_f$ & 2 & 3 &  4-5 & 2 & 3 & 4  \\ \hline 
decouple at $\widehat{A}$ & 
$\tr X^2$ & $\tr X^2$  &   &
$\tr X^2$ & $\tr X^2$  &  \\
decouple at ${A}_k$& 
$M_1,M_2,M_3$ & $M_1$ &  &
$M_{1}, M_2$ &  $M_1$ &  \\ \hline
\end{tabular}
\caption{\label{tab:sqcdtable} For  small values of $(N,k)$, we list all the values of $N_f$ for which $W_{A_k}$ is a relevant deformation of the theory without superpotential, within the regime of vacuum stability $N_f>N/k$. For each $N_f$ in this range, we list all operators that decouple at each of the $\widehat{A}$ and $W_{A_k}$ fixed points.}
\end{table}
\def\arraystretch{1}

Before concluding this section, let us comment on a subtle detail with regards to the final bullet point above. For $N_f < N_f^{\text{decouple}}$, it might be the case that some of the operators $\CO_i$ which generate the chiral ring of the $W_{A_k}$ SCFT -- and which do not themselves decouple at the   $W_{A_k}$ fixed point -- actually decouple at the $\widehat{A}$ fixed point before adding the relevant superpotential deformation. In this case, deforming from the $\widehat{A}$ SCFT by $W_{A_k}$ drives the theory to an interacting fixed point plus free sector which we would {\it not} identify as the $W_{A_k}$ SCFT. This phenomenon occurs in the examples listed in Table \ref{tab:sqcdtable} for which $\tr X^2$ decouples at the $\widehat{A}$ fixed point. In these cases, the $\widehat{A}$ flow does not provide a UV completion of the $W_{A_k}$ fixed point with its full spectrum intact.

\section{A new duality proposal for the \texorpdfstring{$W_{A_k}$}{WAk} SCFTs}
\label{sec:dual}

In this section we introduce the $\CD_{p}[G=SU(N)]$ Argyres-Douglas theories that participate in the proposed duality. We describe their gauging by $\CN=1$ vector multiplets, and examine the conditions for a renormalization group flow to the dual fixed point upon appropriate superpotential deformation.

\subsection{$\mathcal{N}=1$ gaugings of the $\mathcal{D}_p[SU(N)]$ Argyres-Douglas theories}

The theories of type $\CD_p[SU(N)]$ are strongly-coupled $\CN=2$ SCFTs with at least $SU(N)$ flavor symmetry \cite{Xie:2012hs, Cecotti:2012jx, Cecotti:2013lda}, and whose salient properties are reviewed in Appendix \ref{sec:Dp}. We restrict throughout this work to the case of $p,N$ coprime, when the flavor symmetry is exactly $SU(N)$ and many formulae simplify.\footnote{~The choice $\text{gcd}(p,N)=1$ also implies that the $\CD_p[SU(N)]$ theory does not admit a dual Lagrangian quiver description \cite{Cecotti:2013lda}.} We furthermore restrict the range of the positive integer $p$ to $2 \leq p < N$, since $p=1$ trivializes the $\CD_p[SU(N)]$ theory, and since we will only expect a possible duality for $p < N$.

\paragraph{Chiral operators} 
The Coulomb branch of the $\CD_p[SU(N)]$ theories with $\text{gcd}(p,N)=1$ is parameterized by scalar primary operators $u_{j,s}$ of dimension
 \cite{Xie:2012hs, Cecotti:2012jx, Cecotti:2013lda}
    \ba{
    \Delta(u_{j,s})
     =     \left[ j -  \frac{N}{p} s \right]_+ + 1\,,
           \label{UVdim1}
    }
where $\left[ x \right]_+ = x$ for $x>0$ and $0$ for $x\leq0$,
and $j = 1,2,\ldots N-1$, $s=1,2,\ldots p-1$. 
We denote the $(p-1)$ Coulomb branch operators of lowest dimension by $u_i$, whose dimensions satisfy
   \ba{
    \label{uidef1}
    \Delta (u_i)
     =     \frac{p+1+i}{p}, ~~~i=0,1,\ldots p-2\,.
    }
The remaining $\frac{1}{2}(p-1)(N-3)$ Coulomb branch operators form towers above each of the lowest-dimension $u_i$. 

Each Coulomb branch multiplet contains a level-two descendent scalar chiral operator, which upon decomposing the $\CN=2$ multiplet into $\CN=1$ components, is the primary operator of an $\CN=1$ chiral multiplet. We denote by $v_i$ these descendants of the $u_i$, whose dimensions satisfy $\Delta(v_i) = \Delta(u_i) + 1$, in terms of $\Delta(u_i)$ given in \eqref{uidef1},

There is also a conserved current multiplet for the $SU(N)$ flavor symmetry, whose primary is the moment map operator $\mu$ with $\Delta(\mu)=2$, and which transforms in the adjoint representation of $SU(N)$.  This operator satisfies the following chiral ring relations \cite{Xie:2016evu,Song:2017oew, Agarwal:2018zqi},
    \ba{
    \tr \mu^j  =     0\ \ \forall j\,,\qquad
    \mu^p \big|_{{\rm adj}}=     0\,,
    \label{chiralringDpG1}
    }
where $\CO \big|_{{\rm adj}}$ denotes the adjoint part of $\CO$.

 \paragraph{$\CN=1$ deformation} One way to deform the $\CD_p[G]$ theories is to gauge their $G$-flavor symmetry. 
One can in general gauge the diagonal $SU(N)$ flavor symmetry of one or multiple $\CD_{p_i}[SU(N)]$ blocks with either an $\CN=2$ vector multiplet (see \cite{Cecotti:2013lda, Closset:2020afy,Buican:2020moo,Kang:2021lic}), or an $\CN=1$ vector multiplet (as was considered in \cite{Kang:2021ccs, Kang:2023pot}), where the latter breaks $\CN=2$ to $\CN=1$ supersymmetry. Here we gauge the $\CD_{p}[SU(N)]$ theory with an $\CN=1$ vector multiplet, accomplished by adding to the Lagrangian the coupling 
    \bea
  \CL\supset  \int d^4 \theta\, V^a_\mu \mathcal{J}^{\mu a}\,, 
    \eea
  where $V^a_\mu$ is the $SU(N)$ vector multiplet and $\mathcal{J}^{\mu a}$ is the conserved current multiplet, and $a$ is an adjoint index.
  
We furthermore will be interested in coupling $N_f$ flavors of $\CN=1$ chiral multiplets $Q\,(\widetilde{Q})$, which transform in the (anti)fundamental representation of the now-gauged $SU(N)$ symmetry. In other words, this is $\CN=1$ $SU(N)$ SQCD coupled to the $\mathcal{D}_p[SU(N)]$ theory, as pictorially depicted in Figure \ref{fig:dpg}. The $\CD_p[SU(N)]$ block contributes to the 1-loop $\beta$-function as $\left(1 - \frac{1}{p}\right)$ amount of adjoint chiral multiplets (which is encoded in the anomaly coefficient $\tr R G G$ with $G$ denoting the generators of $SU(N)$ symmetry), leading to the following condition for asymptotic freedom of the resulting gauge theory,
\ba{
\label{af}
N_f \leq N\left( 2 + \frac{1}{p} \right)\,.
}

\usetikzlibrary{shapes.geometric}
\usetikzlibrary{arrows, decorations.markings}
		
\begin{figure}[t!]
\centering
	
 		\begin{tikzpicture}[square/.style={regular polygon,regular polygon sides=4}]

\tikzset{
  ->-/.style={decoration={markings, mark=at position 0.5 with {\arrow{stealth}}},
              postaction={decorate}},
}
    
        \draw  (0,0)  ellipse (1.1cm and 0.65 cm);
        \node (a) at (0,0) {$\CD_p[SU(N)]$};


         \draw (1.1,0) -- (2.32,0);

    \begin{scope}[xshift=3.2cm]
	\node[circle,draw=black,minimum size=1.76cm,fill=none]  (1) at (0,0) { };
	\node at (0,0) {$N$};

        \node[square,draw=black,minimum size=2.3cm,fill=none]  (2) at (2.75,0) { };
	\node at (2.75,0) {$N_f$};

 	\draw[->-] (1.north east) -- (2.18-0.25,0.627);
	\draw[->-] (2.18-0.25,-0.627) -- (1.south east);
	\node at (1.3,0.9) {${Q}$};
	\node at (1.3,-0.95) {$\widetilde{Q}$};
    \end{scope}
	\end{tikzpicture}
 \caption{Coupling the $\CD_p[SU(N)]$ theory to $\CN=1$ SQCD. \label{fig:dpg}}
 \end{figure}
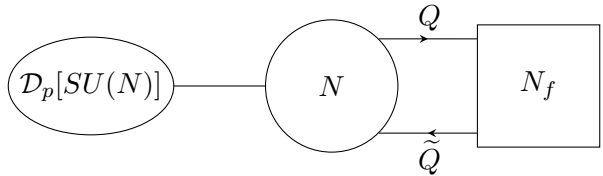

\paragraph{Flow to $\CT_0$}
Within the range \eqref{af}, one can ask if the system of Figure \ref{fig:dpg} flows to an $\CN=1$ SCFT at low energies, which we will denote by $\CT_0$. We will first study this fixed point, before adding a further deformation to flow to the proposed dual of the $W_{A_k}$ SCFTs.

The answer to whether or not there is an interacting IR fixed point will depend on the choices of integers $p,N$, and $N_f$, where recall that we have restricted to $p,N$ co-prime, and $2 \leq p  < N$.  
If the answer is affirmative, then the superconformal R-symmetry at the fixed point is determined by $a$-maximization as a linear combination of the R-symmetry $R_0$ associated to the $\CN=1$ subalgebra of the original $\CN=2$ $U(1)_r\times SU(2)_R$ R-symmetry, given in \eqref{r0}, and a flavor symmetry $F$ given in \eqref{f0}, as
\ba{
R(\epsilon) = R_0 + \epsilon F\,,\qquad R_0 = \frac{1}{3} r + \frac{4}{3} I_3\,,\qquad F = - r + 2 I_3\,.
}
See Appendix \ref{sec:scft} for our supersymmetry conventions, and section \ref{sec:amax} for a review of $a$-maximization in the context of adjoint SQCD.
The R-charges of the chiral fields at the putative fixed point are correspondingly given in terms of $\epsilon$ as,\footnote{~Equivalently, the dimensions $\Delta_{\mathcal{N}=1}(\CO) = 3/2\, R(\CO)$ at the fixed point can  be written as  \bea
    \Delta_{\CN=1}(u_{i})
     =     (1 - 3 \epsilon) \Delta(u_{i}), ~~~
    \Delta_{\CN=1}(v_{i})
     =     1 + 6 \epsilon + (1 - 3 \epsilon) \Delta(u_{i}), ~~
    \Delta_{\CN=1}(\mu)
     =     2 + 3 \epsilon\,,
    \eea
  in terms of $\Delta(\CO)$ the dimension of $\CO$ in the $\CD_p[SU(N)]$ theory before gauging.}
\ba{
\bs{
R(u_i) = \frac{2 (1 + p + i)(1-3\epsilon)}{3p}\,,\qquad R(v_i) = \frac{2 ( 1 + i + 2p - 3 \epsilon (1 + i - p))}{3p} \\
R(\mu) = \frac{4}{3} + 2\epsilon \,,\qquad R(Q) =R(\widetilde{Q})=   1 -\frac{N}{N_f} \frac{(1+2p+3\epsilon(p-1) )}{3p} \,.
}
\label{opdim}}

The gauge-invariant chiral operators formed from these fields include: the scalars $u_i$, $v_i$ which derive from the Coulomb branch multiplets of the $\CD_p[SU(N)]$ theory; traces of products of the moment map operators $\tr \mu^j$; mesonic operators of the form $M_j = Q \mu^{j-1} \widetilde{Q}$; as well as (anti)baryonic operators formed from $N$ (anti)quarks $Q\,(\widetilde{Q})$ and powers of $\mu$, which are analogous to the baryonic operators \eqref{giop} in adjoint SQCD.  However, due to \eqref{chiralringDpG1} the operators $\tr \mu^j$ are absent from the chiral ring, and the mesonic and baryonic operators involving powers of $\mu^{j-1}$ are truncated at $j=p$. The R-charges of the remaining chiral ring operators are,\footnote{~In this table we have listed only the subset $u_i$ of the Coulomb branch operators present in the theory.}
\def\arraystretch{1.35}
\ba{
\begin{array}{|c|c|}
\hline
\CO & R(\CO) \\ \hline
u_i,\  i = 0,\dots,p-2 &  \frac{2 (1 + p + i)(1-3\epsilon)}{3p}\\ 
v_i,\  i = 0,\dots,p-2 & \frac{2 ( 1 + i + 2p - 3 \epsilon (1 + i - p))}{3p} \\ 
M_j = Q \mu^{j-1} \widetilde{Q},\  j = 1,\dots,p  & 2\left(1 - \frac{N}{N_f}\frac{ (1+2p+3\epsilon(p-1) )}{3p}\right) + (j-1)\left(\frac{4}{3} + 2\epsilon\right)   \\
B^{(n_i,\dots ,n_p)} & N\left(1 - \frac{N}{N_f}\frac{ (1+2p+3\epsilon(p-1) )}{3p}\right) + \sum_{\ell = 1}^p n_\ell (\ell - 1) \left(\frac{4}{3} + 2\epsilon\right) \\ \hline
\end{array}
\label{giopsAD}
}
\def\arraystretch{1}

The  trial central charges of the $\CT_0$ theory can be computed as follows. 
Using the 't Hooft anomaly coefficients from Appendix \ref{sec:Dp}, the contributions from the $\CD_p[SU(N)]$ block are,
\ba{
a(\epsilon)_{\CD_p[SU(N)]} &= \frac{ (N^2-1) (1-p)(1 - 3 \epsilon) (1 + 3 \epsilon (1 + 3 \epsilon) - p(2 + 3 \epsilon)^2 ) }{48 p}\,,\\
c(\epsilon)_{\CD_p[SU(N)]} &= \frac{ (N^2-1) (1-p)(1 - 3 \epsilon) ( 3 \epsilon (1 + 3 \epsilon) - p(2 + 3 \epsilon)^2 ) }{48 p}\,,
}
where evidently taking $\epsilon = 0$ reproduces the $\CD_p[SU(N)]$ central charges \eqref{acdpg}. We add to these the contributions from the $\CN=1$ vector multiplet,
\ba{
a_{\CN=1} = \frac{3}{16} (N^2-1)\,,\qquad c_{\CN=1} = \frac{1}{8} (N^2-1)\,,
 }
 as well as the contributions from the quarks $Q,\widetilde{Q}$, 
 \ba{
a(\epsilon)_{Q} = \frac{3}{32} 2 N N_f \left( 3 R(Q)-1)^3 - (R(Q)-1)  \right) \,.
 }
 Performing $a$-maximization on the total,
\ba{a(\epsilon)_{\CT_0}=a(\epsilon)_{\CD_p[SU(N)]}+a_{\CN=1} + a(\epsilon)_{Q}\,,
 }
 yields the following expression for the trial $\epsilon$,
 \ba{
 \label{epst0}
\epsilon = \tfrac{p^3 + N^2 \left( N^2/N_f^2 (1+p-2p^2) - p^3 \right)  + p \sqrt{ p^2 (N^2-1)(N^2 (p^2+p-1)-p^2) +  N^4/N_f^2 (1-p)(N^2 (2+p)-1-2p) }}{3(1-p)( N^4/N_f^2 (1-p) - (N^2-1) p^2)} \,.
 }
 A representative plot of $\epsilon$ as a function of $N_f$ appears in Figure \ref{fig:epsilon}.
Substituting this expression for $\epsilon$ into \eqref{opdim} yields the trial R-charges for the operators \eqref{giopsAD}.
The trial R-charges must be corrected when any operator hits the unitarity bound and becomes free, as can be determined on a case by case basis. 

The range of $N_f$ as a function of $N$ and $p$ for which the flow to the $\CT_0$ fixed point occurs is restricted by considerations analogous to those discussed in section \ref{sec:amax} in the context of the $\widehat{A}$ fixed point. In particular:

	\begin{figure}[t!]
	\centering
 \includegraphics[width=0.55\textwidth]{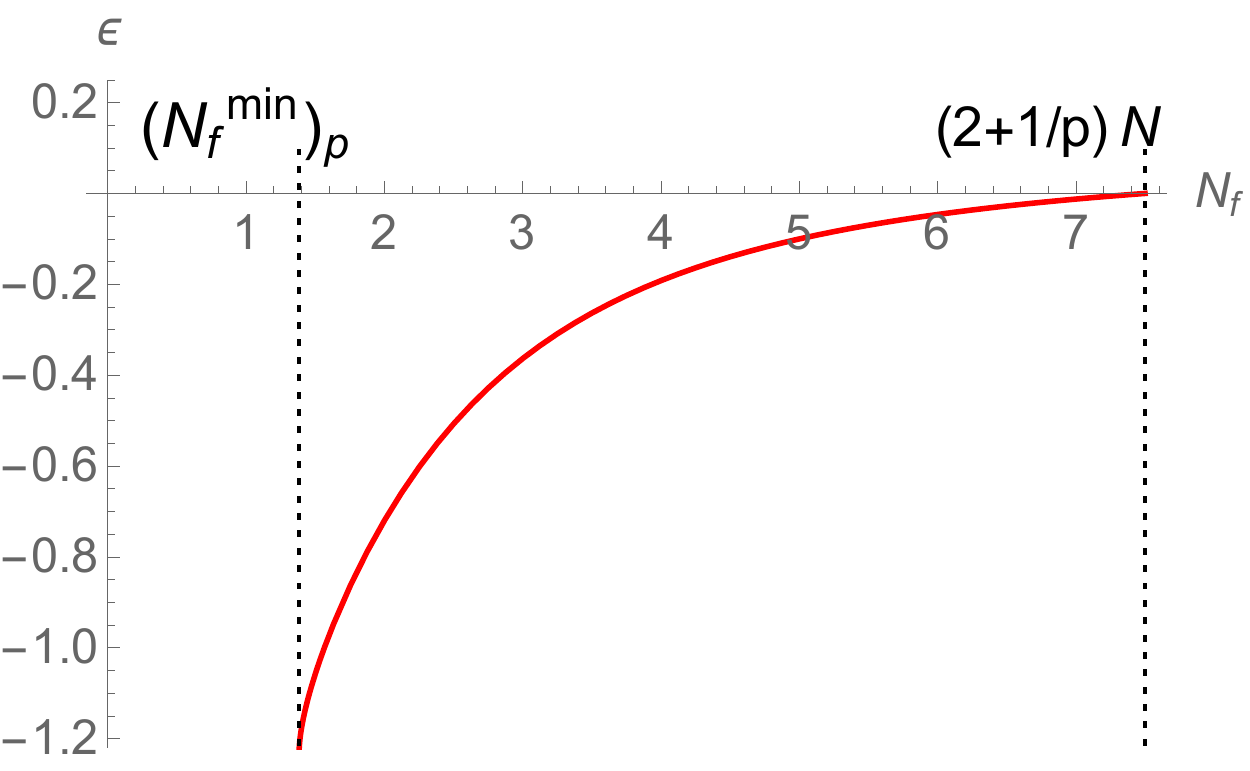}
	\caption{A plot of the trial $\epsilon$ in \eqref{epst0} as a function of $N_f$ for $N=3$ and $p=2$, depicted within the asymptotic freedom bound (on the right) and bound \eqref{nfapprox} (on the left). \label{fig:epsilon}}
	\end{figure}

\begin{itemize}

\item $N_f \leq (2 + 1/p) N$: $N_f$ is bounded from above by the condition of asymptotic freedom of the gauge theory, \eqref{af}. In the limit that this bound is saturated, $\epsilon=0$. 

\item   $N_f > (N_f^{\text{min}})_p$: $N_f$ is bounded from below by the requirement that $\epsilon$ in \eqref{epst0} is non-imaginary, {\it i.e.} that the square root in \eqref{epst0} is nonnegative. 
This results in a bound 
\ba{
 N_f >   (N_f^{\text{min}})_p =  \frac{N^2\sqrt{(p-1)(N^2 (p+2) -2p-1 )}}{ p\sqrt{ (N^2-1)(N^2 (p^2+p-1) - p^2 )}} \approx \frac{N}{p}\,.
 \label{nfapprox}
}
The assertion that this bound is approximated by $\frac{N}{p}$ follows from the fact that for all values $p$ and $N$, $(N_f^{\text{min}})_p$ given in \eqref{nfapprox} lies between
\ba{
\frac{2}{\sqrt{5}} \frac{N}{p} < (N_f^{\text{min}})_p \leq   \frac{N}{p}\,.
}
The quantity $\epsilon$ is negative within the range $(N_f^{\text{min}})_p < N_f  < (2 + 1/p)N$, as depicted in Figure \ref{fig:epsilon} for representative values $N=3$ and $p=2$.   
\item   $N_f > N_f^{\text{decouple}}$: There is also a lower bound on $N_f$ below which operators decouple along the flow to the $\CT_0$ fixed point, which can be more restrictive than the previous bullet point. Generically the meson $M_1=Q\widetilde{Q}$ is the first to hit this bound, at
\ba{
N_f^{\text{decouple}} = \frac{N}{p}\frac{1 + p + \sqrt{ N^2(3 - 3p + p^2) - p^2}}{2 \sqrt{N^2-1}}\,.
}
Below this value of $N_f$, the decoupling of operators that become free must be taken into account in determining the R-symmetry at the fixed point. 

\end{itemize}

\def\arraystretch{1.2}
\begin{table}[t!]
\centering
\begin{tabular}{|c||c||c|c|c||c||} 
\hline
$(N,p)$ & $(3,2)$ &\multicolumn{3}{c||}{$(4,3)$ } & $(5,2)$  \\ \hline
$N_f$ & 4-7 & 2 & 3 & 4-9  & 6-12  \\ \hline 
decouple at $\CT_0$ & & $M_1,M_2$  & $M_1$ &  &  \\
decouple at $W_{u_0}$ &  & $M_1,M_2$ & $M_1$ & & \\ \hline
\end{tabular}

\bigskip

\begin{tabular}{|c||c|c|c|c||c|c|c|c||} 
\hline
$(N,p)$ &\multicolumn{4}{c||}{$(5,3)$ } &\multicolumn{4}{c||}{$(5,4)$ } \\ \hline
$N_f$ & 2 & 3 & 4 & 5-11  & 2 & 3 & 4 & 5-11  \\ \hline 
decouple at $\CT_0$ & $M_1,M_2,M_3,v_0$ & $M_1$ & $M_1$ &  & $M_1,M_2,M_3,v_0$ & $M_1$ & $M_1$ &  \\
decouple at $W_{u_0}$ & $M_1,M_2,M_3$ & $M_1$ & & & $M_1,M_2$ & $M_1$ & &   \\ \hline
\end{tabular}
\caption{\label{tab:dpgtable} For some small values of $(N,p)$, we list all the values of $N_f$ for which the $\CT_0\to W_{u_0}$ flow occurs, including which operators decouple at each fixed point.}
\end{table}
\def\arraystretch{1}

\subsection{Flow from \texorpdfstring{$\CT_0\to W_{u_0}$}{T0 to Wu0}}

We next add the superpotential 
\ba{
\label{supad}
W = u_0\,,
}
where $u_0$ denotes the multiplet that includes the lowest dimension Coulomb branch operator of dimension $\Delta(u_0) = \frac{p+1}{p}$ (see \eqref{uidef1}) and scalar superpartner $v_0$. 
The first task is to determine the conditions for the superpotential  \eqref{supad} to drive a flow to a new $W_{u_0}$ fixed point.
It turns out that as the number of flavors $N_f$ gets smaller, the dimension of the Coulomb branch operators $u_i$ of the $\CD_p[SU(N)]$ theory grows.
Therefore, we have a further restriction on the lower bound in \eqref{nfapprox} from requiring that $u_0$ is a relevant operator at the fixed point. This bound can be determined on a case by case basis; if the trial R-charges do not need any correcting, then substituting \eqref{epst0} into \eqref{opdim} yields a lower bound on $N_f$ as a function of $N$ and $p$, whereas if operators decouple at the $\CT_0$ fixed point, this bound may be pushed to lower values of $N_f$. In Table \ref{tab:dpgtable} we give the allowed values of $N_f$ for small values of $N$ and $p$ that will result in a flow to the $W_{u_0}$ fixed point, including listing the decoupling of operators at the $\CT_0$ fixed point.

When $u_0$ is relevant, the deformation \eqref{supad} by $W=u_0$ initiates an RG flow to the $W_{u_0}$ IR fixed point, and the superconformal $U(1)_R$ symmetry is modified such that the superpotential term has charge two. 
This fixes  $\epsilon$ to the value
    \bea
    \label{epswu}
    \epsilon
     =     \frac{1 - 2p}{3(p+1)}\,,
    \eea
  and the central charges to
    \bea
    \label{acad}
    a
    &=&    \frac{3 \left(-12 N^4+N^2 N_f^2 \left(5 p^2+p+2\right)+N_f^2 \left(-4 p^2+p-1\right)\right)}{8 N_f^2 (p+1)^3}\,, 
            \\
    c
    &=&    \frac{-36 N^4+N^2 N_f^2 \left(16 p^2+5 p+7\right)+N_f^2 \left(-11 p^2+5 p-2\right)}{8 N_f^2 (p+1)^3}\,.
    \eea

  A subset of the gauge-invariant operators that are classically in the chiral ring of the theory and not removed by the ring relation \eqref{chiralringDpG1} are listed in \eqref{giopsAD}. Substituting for $\epsilon$ in \eqref{epswu} yields the following R-charges of these operators,
  \def\arraystretch{1.4}
  \ba{
\begin{array}{|c|c|}
\hline
\CO & R(\CO) \\ \hline
u_i,\  i = 0,\dots,p-2 & \frac{2(p+1+i)}{p+1}\\ 
v_i,\  i = 0,\dots,p-2 & \frac{2 (2+i) }{p+1} \\ 
M_j = Q \mu^{j-1} \widetilde{Q},\  j = 1,\dots,p  &  2\left(1 - \frac{N}{N_f}\frac{2}{1+p} \right)+ (j-1) \frac{2}{1+p} \\
B^{(n_i,\dots ,n_p)} & N\left(1 - \frac{N}{N_f}\frac{2}{1+p} \right) + \sum_{\ell=1}^p n_\ell (\ell-1) \left(  \frac{2}{1+p}\right) \\ \hline
\end{array}
\label{giopsAD2}
}\def\arraystretch{1}
  with scaling dimensions given by $\Delta_{\CN=1} = \frac{3}{2} R$.

For given $p$ and $N$, one can check which operators decouple in the flow to the $W_{u_0}$ fixed point, and revise the central charges \eqref{acad} accordingly. The decouplings at the $W_{u_0}$ fixed point for small values of $N$ and $p$ are given in the bottom lines of Table \ref{tab:dpgtable}.

\subsection{Duality checks}

We propose that the $W_{u_0}$ fixed point obtained from the $\CN=1$ deformed $\CD_p[SU(N)]$ theory depicted in Figure \ref{fig:dpg} by the superpotential deformation \eqref{supad} is dual to the $W_{A_k}$ fixed point of adjoint SQCD with $p=k$.
In this subsection we describe a series of checks of this proposal.\footnote{~It was found in \cite{Xie:2021omd} that the $W_{u_0}$ deformation of the stand-alone $\CD_p[G]$ flows to $|G|$ amount of free chiral multiplets. In our case, when $\CD_p[G]$ is coupled to $G$-gauge fields, it flows to the adjoint chiral multiplet with superpotential $W_{A_p}$.} 

\paragraph{Operator matching}

The matching of the gauge-invariant operators on either side of the duality proceeds by comparing Eq.'s \eqref{giop} and \eqref{giopsAD2}, leading to,
\ba{
\bs{
\tr X^j\quad&\leftrightarrow\quad v_{j-2}\,,\qquad\quad \ \ j=2,\dots,k \\
M_j\quad &\leftrightarrow\quad Q \mu^{j-1} \widetilde{Q}\,,\qquad j=1,\dots,k \\
B^{(n_1,\dots,n_k)} \quad&\leftrightarrow\quad B^{(n_1,\dots,n_k)}
}}
The $k-1$ ring generators made from products of $X$ match one-to-one with the $k-1$ descendants $v_j$ of the Coulomb branch operators in the Argyres-Douglas dual. $X$ and $\mu$ share the same R-charge, as do the quarks on either side of the duality, leading to the matching of the $kN_f^2$ mesons, and baryons. The fact that the mesons and baryons match one-to-one with no superfluous operators is due to the Higgs branch relations \eqref{chiralringDpG1} on the Argyres-Douglas side of the duality. 

In order for the duality to hold, it must be the case that the superfluous Coulomb branch operators $u_{i}$ for $i=1,\dots,k-2$, as well as the additional $(k-1)(N-3)$ Coulomb branch and descendant operators which sit in towers above the $u_i$, are removed from the chiral ring. (The operator $u_0$ is already removed due to the superpotential.) For the special case of $N=3$ and $k=2$, no such additional truncation is required, and the chiral operators evidently match exactly between the two sides of the duality. This case is studied in more detail in section \ref{sec:example} below.
For more general $N$ and $k$, this suggests a quantum chiral ring relation in the $\CN=1$ gauged Argyres-Douglas theory which is not visible at the classical level; such relations can sometimes be detected through the superconformal index, {\it e.g.} see  \cite{Maruyoshi:2018nod}. It would be very interesting to understand the origin of this proposed truncation. 

We note that in the special case with $p=2$ and $N_f = 2N$ this truncation does not occur. Indeed, the fixed point of this case has $\CN=2$ supersymmetry, and in this case all the Coulomb branch operators of the $\CD_2[SU(N)]$ theory are needed to explain the matter content. We will later discuss this issue in section \ref{sec:susyenhance}.

\paragraph{Central charges and 't Hooft anomalies}

The 't Hooft anomalies match between the two sides of the proposed duality, by virtue of the fact that the $\CD_k[SU(N)]$ theory contributes to the 't Hooft anomalies as an $\CN=1$ chiral adjoint multiplet with R-charge $\frac{2}{k+1}$. The $a$ and $c$ central charges in each case are given by \eqref{acak} and \eqref{acad}, and evidently match upon identifying $p=k$.

 \begin{figure}[t!]
 \centering
 \begin{tikzpicture}

 	\draw[->] (-0.75,0) -- (-0.75,6.5);
 	\draw (-0.85,0) -- (0.75,0) node[right] {1} ;
 	\draw (-0.85,1) -- (0.75,1) node[right] {${N}/{k}$} ;
   	\draw (-0.85,1.5) -- (0.75,1.5) node[right] {$\widetilde{W}_{A_k}$ relevant} ;
 	\draw (-0.85,2) -- (0.75,2) node[right] {$W_{u_0}$ relevant} ;
 	\draw (-0.85,4) -- (0.75,4) node[right] {$W_{A_k}$ relevant} ;
  	\draw (-0.85,5) -- (0.75,5) node[right] {$2N$} ;
     \draw (-0.85,6) -- (0.75,6) node[right] {$2N + {N}/{k}$} ;
 	\node at (0,6.8) {$N_f$};

 	\fill[fill=red,opacity=0.1] (0.25,1) -- (0.75,1) -- (0.75,4) -- (0.25,4);
 	\fill[fill=green,opacity=0.1] (-0.25,2) -- (-0.75,2) -- (-0.75,6) -- (-0.25,6);
   	\fill[fill=blue,opacity=0.1] (-0.25,1.5) -- (0.25,1.5) -- (0.25,4) -- (-0.25,4);
	
 	\draw[fill=red,opacity=0.1] (5,3.5) -- (11,3.5) -- (11,4) -- (5,4);
 	\node at (8,3.75) {Adjoint SQCD flows to fixed point};
 	\draw[fill=blue,opacity=0.1] (5,2.5) -- (11,2.5) -- (11,3) -- (5,3);
     	\node at (7.8,2.75) {KS dual conformal window};
       	\draw[fill=green,opacity=0.1] (5,1.5) -- (11.2,1.5) -- (11.2,2) -- (5,2);
 	\node at (8.1,1.75) { $\CD_k[SU(N)]$ dual flows to fixed point};

 \end{tikzpicture}
 \caption{A schematic representation of the conformal windows for dual descriptions of the $W_{A_k}$ SCFTs. The overlap of the green and red regions corresponds to the conformal window of the proposed duality with the $\CN=1$ deformed $\CD_k[SU(N)]$ theory. Here $\widetilde{W}$ refers to the superpotential for the Kutasov-Schwimmer (KS) dual theory. \label{fig:conformalwindow}}
 \end{figure}
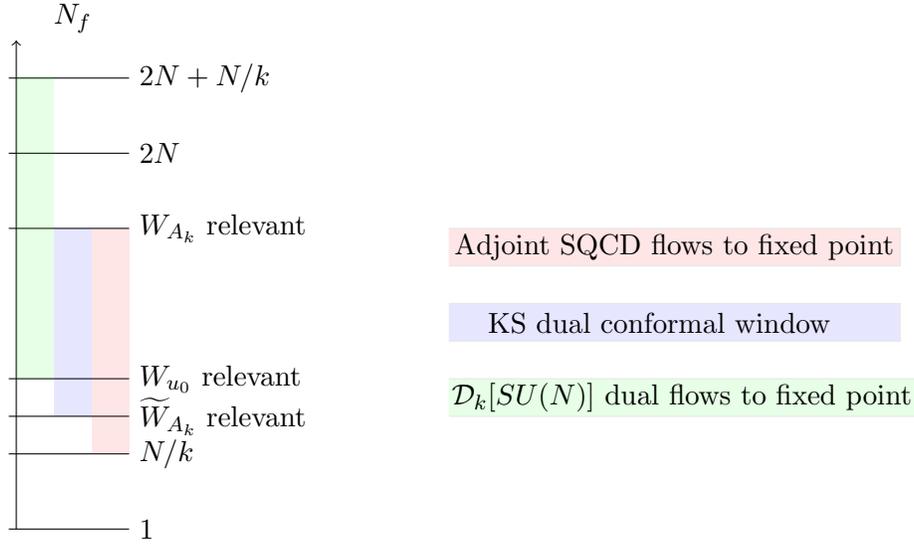

\paragraph{Conformal window}

The conformal window for the proposed dual theories is summarized by Figure \ref{fig:conformalwindow}. 
As discussed in section \ref{sec:dual}, the bounds on  $N_f$ as a function of $N$ and $p$ for the $\CN=1$ deformed Argyres-Douglas gauge theory to flow to the $W_{u_0}$ fixed point lie within the range $N/p \lesssim N_f \leq (2 + 1/p)N$, with this range further restricted from below by checking when $u_0$ is a relevant deformation of the theory without superpotential. This range can be compared with that discussed in section \ref{sec:Ak} for the flow from adjoint SQCD to the $W_{A_p}$ fixed point, which occurs within the window $N/p < N_f \leq 2N$, and is further restricted from above by checking when $\tr X^{p+1}$ is a relevant operator in the $\widehat{A}$ theory without superpotential. For all values of $N$ and $p$ that we checked, there is a nontrivial conformal window for which these two regions overlap. For low values of $N$ these are given by the overlap of Tables \ref{tab:sqcdtable} and \ref{tab:dpgtable}; for instance, 
\ba{
\label{cwad}
&\begin{array}{cc|c}
& & \text{conformal window} \\ \hline \hline
N=3 & k=2 &  4 \leq N_f \leq 6  \\ \hline
N=4 & k=3 & 2 \leq N_f \leq 4 \\ \hline 
N=5 & k=2 & 6 \leq N_f \leq 10
\end{array} 
}

A few comments are in order regarding the range of the proposed duality.

\begin{itemize}

\item Interestingly, the vacuum stability bound $N_f > N/k$ of the $W_{A_k}$ fixed point approximately matches onto the necessary bound $N_f \gtrsim N/k$ derived in \eqref{nfapprox} for the dual Argyres-Douglas theory to flow to the $\CT_0$ fixed point.

\item In the flow to the $W_{A_k}$ fixed point from adjoint SQCD, the requirement that the superpotential be relevant lowered the value of $N_f^{\text{max}}$, whereas in the Argyres-Douglas dual this requirement raised the value of $N_f^{\text{min}}$. It is precisely these factors that are most restrictive in carving out the conformal window. This is similar to the scenario in the  Kutasov-Schwimmer dual, where the requirement that the magnetic superpotential $\widetilde{W}_{A_k}$ is relevant raises the bottom of the conformal window (see Figure \ref{fig:conformalwindow}).

\item However, we emphasize that unlike the case with the magnetic dual, this is {\it not} a strong-weak coupling type of duality in the usual sense. In both the electric description of the $W_{A_k}$ fixed point and the proposed Argyres-Douglas dual, increasing the number of flavors brings one closer to the asymptotic freedom bound of the 1-loop $\beta$-function. 

\item One subtlety in determining the conformal window comes from the decoupling of operators at the $\CT_0$ fixed point. As in the comments below \eqref{nfdec}, it is sometimes the case that more operators decouple at the $\CT_0$ fixed point than at the $W_{u_0}$ fixed point. Of the examples in Table \ref{tab:dpgtable}, this is the case {\it e.g.} for $(N,k)=(5,3)$ and $(N,k)=(5,4)$ with $N_f=2$, since for these values $v_0\sim \tr X^2$ decouples at $\CT_0$. In such cases, we would say that the RG flow that begins at $\CT_0$ does not provide a UV completion of the $W_{A_k}$ fixed point with all its operator spectrum intact. This is analogous to the discussion in section \ref{sec:amax} regarding the flow from $\widehat{A}\to W_{A_k}$, where this phenomenon also occurs. 

\end{itemize}

\paragraph{Superconformal index} 

The superconformal index  of an $\CN=1$ superconformal theory is defined as \cite{Romelsberger:2005eg, Kinney:2005ej}
\begin{align} \label{eq:N1idx}
\begin{split} 
    I &= \Tr (-1)^F t^{3(R+2j_2)} y^{2j_1} \prod_i a_i^{F_i} \\
     &= \Tr (-1)^F p^{j_1 + j_2 + R/2} q^{j_2 - j_1 + R/2} \prod_i a_i^{F_i} \, .
\end{split}
\end{align}
The trace is taken over the states satisfying $\Delta = \frac{3}{2}R + 2j_2$ where $(\Delta, j_1, j_2, R)$ are the Cartans of the bosonic subgroup of the superconformal group, denoting the scaling dimension, Lorentz spins, and R-charge, respectively. $F_i$ denotes the Cartans for the flavor symmetry generators. The second line is obtained via mapping the fugacities as $p=t^3 y, q=t^3/y$. These two forms are used interchangeably. 
We can compute the full superconformal index of the $\mathcal{D}_p[SU(N)]$ theory for $N=2$ and for all $p$, using the $\CN=1$ Lagrangian of \cite{Maruyoshi:2016aim, Maruyoshi:2016tqk, Agarwal:2016pjo}. Other than $N=2$, only the index of the $\mathcal{D}_2 [SU(3)]$ theory is known. We will consider the full indices for these low rank gauge groups separately. 

Even though the full index for the general $\mathcal{D}_p[SU(N)]$ theory is not known, a special limit of the index known as the Schur index is known to have a particularly illuminating closed form when $p$ and $N$ are coprime. To define the Schur index, let us first start with a general $\CN=2$ superconformal index, which takes the form
\begin{align} \label{eq:N2idx}
    I_{\CN=2}(p, q, \mathfrak{t}) = \Tr (-1)^F p^{j_1+j_2+r/2} q^{j_2 - j_1 + r/2} \mathfrak{t}^{I_3 - \frac{1}{2} r} \, , 
\end{align}
where $I_3$ and $r$ denote the Cartan of the $SU(2)_R$ and the generator of $U(1)_r$ respectively. 
The Schur limit is defined as taking $q = \mathfrak{t}$. In this limit, it turns out that the $p$-dependence drops out \cite{Gadde:2013fma}. 

The Schur index of the $\mathcal{D}_p[SU(N)]$ theory is given as \cite{Xie:2016evu, Song:2017oew}
\begin{align} \label{eq:DpGSchurIdx}
    I_S^{\mathcal{D}_p[SU(N)]} = \textrm{PE} \left[ \frac{q - q^p}{(1-q)(1-q^p)} \chi_{\textrm{adj}}(z) \right] \, , 
\end{align}
where PE stands for the plethystic exponential $\textrm{PE}[z]\equiv \exp(\sum_{n\ge 1} \frac{z^n}{n})$ and $\chi_{\textrm{adj}}$ denotes the character of the adjoint representation of $SU(N)$. This form is particularly interesting since it can be compared with that of an 
$\CN=1$ chiral multiplet, which is given as
\begin{align} \label{eq:ChiralSchurIdx}
    I_{\textrm{chiral}} (p, q; a) = \textrm{PE}\left[ \frac{(pq)^{r/2} a - (pq)^{1-r/2} a^{-1}}{(1-p)(1-q)}\right] \, , 
\end{align}
where $r$ is the $R$-charge of a chiral multiplet in the IR and $a$ denotes the fugacity for the flavor (or gauge, if it is couped to a gauge field) symmetry acting on a chiral multiplet. Notice that if we take $p=q^p$ and also $r=\frac{2}{p+1}$ in \eqref{eq:ChiralSchurIdx}, it becomes identical form to that of \eqref{eq:DpGSchurIdx}:
\begin{align}
    I_{\textrm{chiral}} (p, q) \Big|_{p = q^p, r = \frac{2}{p+1}} = \textrm{PE} \left[ \frac{q a - q^p a^{-1}}{(1-q)(1-q^p)} \right]\,.
\end{align}
Therefore, if we have a gauge theory with a chiral multiplet of $R$-charge $\frac{2}{p+1}$ in the adjoint representation of $SU(N)$, in the limit $p \to q^p$, the ($\CN=1$) superconformal index becomes  identical to the Schur index \eqref{eq:DpGSchurIdx} of the $\mathcal{D}_p[SU(N)]$ theory. This can be indeed realized by SQCD with an adjoint $X$ and superpotential $W = \tr X^{p+1}$. 

The superconformal index for SQCD coupled to the $\CD_p [SU(N)]$ theory (see Figure \ref{fig:dpg}) can be schematically written as
\begin{align}
    I = \int [dz] I_{\textrm{vec}}(z) I_{\textrm{chi}}(z) I_{\CD_p[SU(N)]}(z)\Big|_{\mathfrak{t} \to (pq)^{\frac{2}{3}+\epsilon}} \ , 
\end{align}
where $I_{\textrm{vec}}$ and $I_{\textrm{chi}}$ denote the index for an $\CN=1$ vector and chiral multiplet respectively. $I_{\CD_p[SU(N)]}$ denotes the ($\CN=2$) index for the $\CD_p[SU(N)]$ theory, and we rewrite $\mathfrak{t}$ appropriately to cast it as an $\CN=1$ index. The mixing parameter $\epsilon$ is fixed via $a$-maximization to give \eqref{epswu} once it is deformed by the lowest-dimensional Coulomb branch operator in $\CD_p[SU(N)]$. 
Since we do not know the full index in general, we would like to take $q = \mathfrak{t}$ limit. This results in 
\begin{align}
    q = (p q)^{\frac{2}{3} + \epsilon}  \quad \to \quad p = q^p \ , 
\end{align}
so that the contribution to the index from the $\CD_p[SU(N)] $ theory upon RG flow via $\CN=1$ gauging and the $W=u_0$ deformation becomes identical to that of  adjoint SQCD. Thus, the superconformal indices on both sides of the proposed duality match exactly in the $p = q^p$ limit.\footnote{~This type of (Schur-like) limit for the $\CN=1$ theory has been considered in \cite{Buican:2016hnq} to analyze $\CN=1$ deformations of the minimal Argyres-Douglas theory.}

As we can see from the form of the index, if the duality is true, we find that the specialization of the full index for the $\CD_p[G]$ theory should agree with that of the free chiral multiplets with $R$-charge $\frac{2}{p+1}$:
  \begin{align} \label{eq:idxDpGtoChiral}
      I_{\CD_p [G]} (z; p, q, \mathfrak{t}) \Big|_{\mathfrak{t} \to (pq)^{\frac{1}{p+1}}} = \prod_{\alpha \in \Delta_G} \Gamma( (pq)^{\frac{1}{2(p+1)}} z^\alpha ) \,.
  \end{align}
Here $z^\alpha \equiv \prod_i z_i^{\alpha_i}$ and $\Delta_G$ is the set of all roots for the $G$. The elliptic Gamma function is given as
\begin{align}
    \Gamma(z) = \prod_{n, m \ge 0} \frac{1- z^{-1} p^{m+1}q^{n+1}}{1-z p^m q^n} \ , 
\end{align}
which gives the index for chiral multiplet. 
The relation \eqref{eq:idxDpGtoChiral} is also consistent with the Schur-like limit ($p \to q^p$) we considered above.

\subsection{Supersymmetry enhancement for \texorpdfstring{$N_f=2N$}{Nf=2N} and $p=2$}
\label{sec:susyenhance}

  Let us now consider the special case of $p=2$ and $N_f = 2N$ for the Argyres-Douglas dual theory. We propose that upon further deformation by the superpotential 
    \bea
    W
     =     Q \mu \widetilde{Q}\,,
     \label{QmuQ}
    \eea
  supersymmetry enhancement occurs at low energies. 
  The central charges at the fixed point are
    \bea
    a
     =     \frac{7N^2 - 5}{24}\,, ~~~
    c
     =     \frac{2N^2 - 1}{6}\,,
    \eea
  which are identical to those of $\CN=2$ SQCD with $N_f=2N$. Furthermore, the chiral operators can be identified with those of $\CN=2$ SQCD as follows: since the dimensions of $Q$, $\widetilde{Q}$ and $\mu$ are all $1$, these give the mesonic and baryonic operators of $\CN=2$ SQCD; the dimensions of $u_{j,1}$ and $v_{j,1}$ (defined around \eqref{UVdim1}) are 
    \bea
    \Delta(u_{j,1})
     =     3, 5, \ldots, N\,, ~~~~
    \Delta({v_{j,1}})
     =     2, 4, \ldots, N-1\,,
    \eea
   which are the same as those of the $\Tr X^k$ operators ($k=2, 3, \ldots, N$) of $\CN=2$ SQCD. Note that this mapping of operators is similar to that of $\CN=4$ enhancement studied in  \cite{Kang:2023pot}.

   Once the superpotential \eqref{QmuQ} is added, the IR theory lands on the interacting $\CN=2$ fixed point (or along the conformal manifold). This is mapped to $W=Q X \widetilde{Q}$ in the adjoint SQCD dual as we will see below, and breaks the global $SU(N_f)\times SU(N_f)$ symmetry to the diagonal part. This is actually crucial since when this coupling is turned off, the gauge coupling is also driven to zero \cite{Leigh:1995ep}. 

   As in the previous subsection, we can check that the superconformal index agrees with that of $\CN=2$ SQCD at least in the limit $p=q^2$. This is clear, since in this case the Schur index of the $\CD_2[SU(N)]$ theory is equal to the index of the adjoint chiral with R-charge $2/3$. Since we do not know the superconformal index of the $\CD_2[SU(N)]$ theory in  full generality it is currently not possible to check beyond this limit, except for $N=3$. In the next subsection we compute the full index for the $N=3$ case explicitly, and verify that it matches.

   A potential issue with our proposal is that the $U(1)_F$ symmetry in the $\mathcal{D}_2[SU(N)]$ theory is broken by the $u_0$ deformation. We propose that the accidental $U(1)_F$ symmetry emerges in the IR to form the $\CN=2$ R-symmetry. This can be understood from the adjoint SQCD description. 
   Let us first consider the adjoint SQCD theory without $\Tr X^{p+1}$ deformation. The matter content is the same as that of $\CN=2$ SQCD with $N_f = 2N$ flavors. With the superpotential $Q X \widetilde{Q}$ which is dual to \eqref{QmuQ}, the theory flows to the fixed point with $\CN=2$ as discussed in \cite{Leigh:1995ep}. At this point, the $\Tr X^{3}$ operator (which is supposed to be mapped to $u_0$ in the Argyres-Douglas dual theory) is marginal, but is irrelevant at low energies since it breaks the global symmetry (namely $U(1)_r$). Therefore the low energy theory of adjoint SQCD has $\CN=2$ supersymmetry.

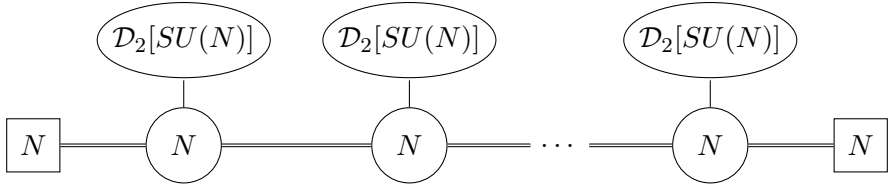
\begin{figure}[t!]
\centering
  \begin{tikzpicture}[square/.style={regular polygon,regular polygon sides=4}]
  \tikzset{
  ->-/.style={decoration={markings, mark=at position 0.5 with {\arrow{stealth}}},
              postaction={decorate}},
  }
  
  \draw  (2,1.4)  ellipse (1.15cm and 0.5 cm);
  \draw  (5,1.4)  ellipse (1.15cm and 0.5 cm);
  \draw  (9,1.4)  ellipse (1.15cm and 0.5 cm);
  \node (a0) at (2,1.4) {$\CD_2[SU(N)]$};
  \node (b0) at (5,1.4) {$\CD_2[SU(N)]$};
  \node (c0) at (9,1.4) {$\CD_2[SU(N)]$};
  \node (a1) at (2,1) { };
  \node (b1) at (5,1) { };
  \node (c1) at (9,1) { };

  \node[square,draw=black,minimum size=1.0cm,fill=none]  (1) at (0,0) { };
  \node at (0,0) {$N$};
 
  \node[circle,draw=black,minimum size=1.0cm,fill=none]  (2) at (2,0) { };
  \node at (2,0) {$N$};
  
  \node[circle,draw=black,minimum size=1.0cm,fill=none]  (3) at (5,0) { };
  \node at (5,0) {$N$};
  
  \node (x) at (7,0) {$\dots$};
  
  \node[circle,draw=black,minimum size=1.0cm,fill=none]  (4) at (9,0) { };
  \node at (9,0) {$N$};
  
  \node[square,draw=black,minimum size=1.0cm,fill=none]  (5) at (11,0) { };
  \node at (11,0) {$N$};

  \draw[double] (1) -- (2);
  \draw[double] (2) -- (3);
  \draw[double] (3) -- (x);
  \draw[double] (x) -- (4);
  \draw[double] (4) -- (5);
  
  \draw (2) -- (a1);
  \draw (3) -- (b1);
  \draw (4) -- (c1);
  
  \end{tikzpicture}
 \caption{Linear quiver gauge theory coupled to $\CD_2[SU(N)]$ theories.
 The double line denotes a pair of bi-fundamental chiral multiplets transforming in the $(\square, \overline{\square})$ and $(\overline{\square}, \square)$ representations of the left and  right $SU(N)$ symmetries. 
 \label{fig:linearquiver} }
 \end{figure}

   It is straightforward to generalize this idea to quiver gauge theories. Consider the linear quiver gauge theory depicted in Figure \ref{fig:linearquiver} where all the gauge groups are $\CN=1$ and $SU(N)$. Each gauge (or flavor) node is connected by a pair of the bi-fundamental chiral multiplets, and is coupled to the $\CD_2[SU(N)]$ theory via the superpotential 
     \bea
     W
      =     \sum_{a=1}^\ell \left( u_{0a} + (\widetilde{Q}_a Q_a - Q_{a+1}\widetilde{Q}_{a+1}) \mu_a \right) ,
    \eea
  where $u_{0a}$ and $\mu_a$ are the chiral operators from the $\CD_2[SU(N)]$ theories coupled to the $a$-th gauge nodes; $Q_a$ and $Q_{a+1}$ are the bi-fundamental chiral multiplets coupled to the $a$-th gauge nodes from the left and the right respectively.
  As in the previous case, the $\CD_2[SU(N)]$ theory plays the role of the adjoint chiral multiplet with R-charge $2/3$ at low energies. Therefore, we expect that the fixed point theory is $\CN=2$ linear quiver gauge theory. This can be checked at the same level as the Argyres-Douglas dual theory discussed throughout this paper. 
   
   One would be able to construct a circular quiver theory by gauging the diagonal symmetry of the left-most and the right-most flavor nodes with an additional $\CD_2[SU(N)]$,  and with the superpotential $W=u_0$ turned on. This is expected to flow to an $\CN=2$ circular quiver gauge theory.

\subsection{Case study: dualities for the \texorpdfstring{$W_{A_2}$}{WA2} SCFT with \texorpdfstring{$N=3$}{N=3}}
\label{sec:example}

  Let us consider the proposed duality in detail for the $W_{A_2}$ SCFT with $N=3$. 
  The number of the flavors is restricted to $3<N_f\leq 6$, as per \eqref{cwad}.

  The dual theory is $\CN=1$ $SU(3)$ gauge theory coupled to $\mathcal{D}_2[SU(3)]$ and $N_f$ flavors.
  The $\mathcal{D}_2[SU(3)]$ theory (which is equivalent to the $(A_1, D_4)$ Argyres-Douglas theory) has a one-dimensional Coulomb branch parameterized by the Coulomb branch operator $u_0$ of dimension $\frac{3}{2}$.
  The associated operator ($\CN=2$ superpartner) $v_0$ has dimension $\frac{5}{2}$. 

\paragraph{$N_f=4$:}
  Let us focus first on the $N_f=4$ case.
  Without any superpotential, the gauge theory flows to the fixed point with the central charges $(a, c) \simeq (2.036, 2.497)$, where $\Delta_{\CT_0}(u_0) \simeq 2.364$.
  After the deformation by $u_0$, the IR fixed point has the central charges 
    \bea
    (a, c)
     =     \left( \frac{187}{96}, \frac{277}{96} \right).
    \eea
  These are equivalent to those of the $W_{A_2}$ SCFT with $N=3$ and $N_f=4$.

  The operator matching perfectly works in this case,
    \bea
    \Tr X^2
    &\leftrightarrow &
           v_0\,, \nonumber \\
    M_j
    &\leftrightarrow &
           Q \mu^{j-1} \widetilde{Q}\,, ~~~  j=1,2 \\
    B^{(n_1,n_2)}
    &\leftrightarrow &
           B^{(n_1,n_2)}\,, \nonumber
    \eea
  with no superfluous operators to account for. 
  The dimensions of these operators at the fixed point are given by
    \bea
    \Delta_{{\rm IR}}(v_0)
     =     2\, , ~~~
    \Delta_{{\rm IR}}(Q \mu^{j-1} \widetilde{Q})
     =     \frac{1}{2} + j\, , ~~~
    \Delta_{{\rm IR}}(B^{(n_1,n_2)})
     =     \frac{9}{4}+n_2\, .
    \eea

  For this case, we can compute the full superconformal index using the $\CN=1$ Lagrangian flowing to $\CD_2[SU(3)] = (A_1, D_4)$ theory \cite{Maruyoshi:2016aim, Maruyoshi:2016tqk, Agarwal:2016pjo}. It can be computed using the integral expression
  \begin{align} \label{eq:idxADdual}
    \begin{split}
      I &= \oint [dz]_G I_{\textrm{vec}}(z) I_{\textrm{chi}}(z) \, I_{\CD_p[SU(N)]}(z) \Big|_{\mathfrak{t}\to (pq)^{\frac{2}{3}+\epsilon}} \\
       &= \kappa^{N-1} \oint \prod_{i=1}^N \frac{dz_i}{2\pi i z_i} \prod_{i \neq j} \frac{1}{\Gamma(z_i/z_j)} \prod_{i, a} \Gamma( (p q)^{\frac{r}{2}} z_i^\pm x_a^\pm ) \, I_{\CD_p[SU(N)]}(z; p, q, \mathfrak{t}) \Big|_{\mathfrak{t}\to (pq)^{\frac{2}{3}+\epsilon}} \ , 
    \end{split}
  \end{align}
  where the gauge indices $i, j$ takes value in $i, j=1, \ldots, N$ under the constraint $\prod_i z_i = 1$, and the flavor index $a$ run from $a= 1, \ldots N_f$. Here $\Gamma(z) \equiv \prod_{m, n \ge 0} \frac{1- z^{-1} p^{m+1}q^{n+1}}{1-z p^m q^n}$ is the elliptic Gamma function and $\kappa \equiv (p; p)(q; q)$ with $q$-Pochhammer symbol defined as $(z; q) \equiv \prod_{n\ge 0} (1-z q^n)$. For the contribution from the $\CD_p[G]$ block, we rewrite the $\CN=2$ fugacity to $\CN=1$ by taking $\mathfrak{t} \to (pq)^{\frac{2}{3}+\epsilon}$ where $\epsilon$ is the R-symmetry mixing parameter given as \eqref{epswu} and $r$ is the $R$-charge for the fundamentals as is given in \eqref{opdim}. 
  Upon evaluating the integral for $N=3, N_f=4$ (which sets $r=1/2, \epsilon = -1/3$) and expanding in $t$ with $p=t^3 y, q=t^3/y$, the (reduced) superconformal index is given as
    \begin{align}
    \begin{split}
    I_{\textrm{red}} &\equiv (I-1)(1-t^3/y)(1-t^3 y) \\  
     &= t^3 \left(1 + \chi_{[1, 0, 1]}(x) \right) + t^4 + t^{\frac{9}{2}} \left(a^3 \chi_{[0, 0, 1]}(x)+ a^{-3}\chi_{[1, 0, 1]}(x) \right) \\
     &~~+ t^5 \left(-\chi_{\frac{1}{2}}(y) + 1+ \chi_{[1, 0, 1]}(x) \right) + t^6 \left(\chi_{[2, 0, 2]}(x)+\chi_{[0, 2, 0]}(x)+1 \right) \\
     &~~+ t^{\frac{13}{2}} \left(a^3 \chi_{[1, 1, 0]}(x)+ a^{-3} \chi_{[0, 1, 1]}(x) \right) + t^7 \left( 1+\chi_{[1, 0, 1]}(x) \right) \\ 
     &~~+ t^{\frac{15}{2}} (a^3 \chi_{[1, 0, 2]}(x) + a^{-3}\chi_{[2, 0, 1]}(x) ) + t^8 \left(-\chi_{\frac{1}{2}}(y)(1+\chi_{[1, 0, 1]}(x) ) \right)\\
     &~~+ t^8 \left( \chi_{[2, 0, 2]}(x) + \chi_{[0, 1, 2]}(x) + \chi_{[2, 1, 0]}(x) + \chi_{[0, 2, 0]}(x) + 2\chi_{[1, 0, 1]}(x) + 2 \right) + \ldots \ , 
    \end{split}
    \end{align}
    where $\chi_j(y)$ denotes the characters for the spin-$j$ irrep of $SU(2)$ subgroup of the Lorentz group and $\chi_{[n_1, n_2, n_3]}(x)$ denotes the character for the representation of flavor $SU(4)$ symmetry with Dynkin label $[n_1, n_2, n_3]$, and $a$ is the fugacity for the baryonic $U(1)$ symmetry rotating the fundamental flavors. 
    
  We check that this is precisely equal to the superconformal index of $SU(3)$ adjoint SQCD with $N_f=4$ deformed by $\Tr X^3$. The index for adjoint SQCD with $W=\Tr X^{p+1}$ can be written as
  \begin{align}
      \begin{split}
       I &= \left(\kappa \Gamma ((pq)^{\frac{1}{p+1}}) \right)^{N-1} \oint \prod_{i=1}^N \frac{dz_i}{2\pi i z_i} \prod_{i \neq j} \frac{\Gamma( (pq)^{\frac{1}{p+1}} z_i / z_j) }{\Gamma(z_i/z_j)} \prod_{i, a} \Gamma( (p q)^{\frac{r}{2}} z_i^\pm x_a^\pm ) \ , 
      \end{split}
  \end{align}
  and it agrees with the index of the dual theory for $N=3, N_f=4$ with $p=2$, as we have verified up to order $t^8$. 
  
\paragraph{$N_f=6$: SUSY Enhancement to $\CN=2$}
  Let us move to the next example with $N_f=6$.
  Without the superpotential, the gauge theory flows to a fixed point with central charges $(a, c) \simeq (2.768, 3.274)$, where $\Delta_{\CT_0}(u_0) \simeq 1.708$.
  The deformation by $u_0$ causes the RG flow to the fixed point with the central charges 
    \bea
    (a, c)
     =     \left( \frac{29}{12}, \frac{17}{6} \right).
    \eea
  One may verify that these coincide with the central charges of  $\CN=2$ $SU(3)$ SQCD with $N_f=6$ flavors.
  We propose that at this fixed point, the supersymmetry is enhanced to $\CN=2$.
  This is quite natural from the dual adjoint SQCD description: the matter content is simply that of $\CN=2$ $SU(3)$ SQCD with $N_f=6$ flavors.
  Without the superpotential the theory flows to the $\CN=2$ fixed point  at zero gauge coupling \cite{Leigh:1995ep}, so we turn on $W=Q\mu\widetilde{Q}$ in order to be at an interacting fixed point. 
  
  Again the issue is that the $U(1)_F$ symmetry in the $\mathcal{D}_2[SU(3)]$ theory is broken by the $u_0$ deformation. Now, $u_0 \leftrightarrow \Tr X^3$ is a deformation on the Coulomb branch, but it is marginally irrelevant since it breaks a flavor symmetry carried by $X$ \cite{Leigh:1995ep, Green:2010da}. Since this deformation is marginally irrelevant, we expect the accidental $U(1)_F$ symmetry to emerge at low energies. 

  The (reduced) superconformal index can be calculated from \eqref{eq:idxADdual} with $p=2,N=3, N_f=6$, for which we obtain
    \begin{align} \label{eq:idxSU3Nf6AD}
    \begin{split}
    I_{\textrm{red}} &\equiv (I-1)(1-t^3 y)(1-t^3/y) \\
    &= t^4 (2 + \chi_{[1, 0, 0, 0, 1]}) - t^5 \chi_{\frac{1}{2}}(y) + t^6 ((a^3 + a^{-3})\chi_{[0, 0, 1, 0,0]} - \chi_{[1, 0, 0, 0, 1]} ) \\
    &~~+ t^8 (- (a^3 + a^{-3})\chi_{[0, 0, 1, 0, 0]} + \chi_{[2, 0, 0, 0, 2]}+\chi_{[0, 1, 0, 1, 0]}+2 \chi_{[1, 0, 0, 0, 1]}) \ldots \ , 
    \end{split}
    \end{align}
  where $a$ is the fugacity of the baryonic $U(1)$ and $\chi_{[n_1, n_2, n_3, n_4, n_5]}$ denotes the character for the flavor $SU(6)$ symmetry.
  
  Now the index of the $\CN=2$ $SU(N)$ SQCD with $N_f=2N$ is given as 
  \begin{align}
       I &= \left(\kappa \Gamma (\frac{pq}{\mathfrak{t}}) \right)^{N-1} \oint \prod_{i=1}^N \frac{dz_i}{2\pi i z_i} \prod_{i \neq j} \frac{\Gamma( (\frac{pq}{\mathfrak{t}}) z_i / z_j) }{\Gamma(z_i/z_j)} \prod_{i, a} \Gamma( \mathfrak{t}^{1/2} z_i^\pm x_a^\pm ) \ , 
  \end{align}
  with $a=1, \ldots 2N$. Upon evaluating the integral, we obtain
  \begin{align} \label{eq:idxSU3Nf6}
  \begin{split}
    I_{\textrm{red}} &\equiv (I-1)(1-t^3 y)(1-t^3/y) \\
    &= t^4 (v^{-1}(1 + \chi_{[1, 0, 0, 0, 1]}) + v^2) - t^5 v \chi_{\frac{1}{2}}(y) \\
    &~~+ t^6 (v^3 - v^{\frac{1}{2}} + v^{-\frac{3}{2}} (a^3 + a^{-3})\chi_{[0, 0, 1, 0,0]} - \chi_{[1, 0, 0, 0, 1]} ) + t^7 \chi_{\frac{1}{2}}(y)(v^{-\frac{1}{2}} - v^2) \\
    &~~+ t^8 (-v^{\frac{3}{2}} + v^4 - v^{-\frac{3}{2}}(\chi_{[1, 0, 0, 0, 1]}+1) + v \chi_{[1, 0, 0, 0, 1]} + 3)  \\ 
    &~~~- v^{-\frac{1}{2}} (a^3 + a^{-3})\chi_{[0, 0, 1, 0, 0]}  + v^{-2} (\chi_{[2, 0, 0, 0, 2]} + \chi_{[0, 1, 0, 1, 0]} +2 \chi_{[1, 0, 0, 0, 1]} + 2) +  \ldots , 
  \end{split}
  \end{align}
  with the fugacity $v$ corresponding to the $U(1)_F$ symmetry, which is a combination of $I_3$ and $r$ (more precisely $r/2 - I_3$ or equivalently $\mathfrak{t} = t^4/v$ in \eqref{eq:N2idx}). Upon specializing to $N=3$, $v=1$, we see that the indices \eqref{eq:idxSU3Nf6AD} and \eqref{eq:idxSU3Nf6} agree. 
  As stated above, the corresponding $U(1)_F$ symmetry cannot be seen from the UV description. 

\section{Conclusions and speculations for the \texorpdfstring{$D$}{D}- and \texorpdfstring{$E$}{E}-series}
\label{sec:conclusion}

We proposed that $\CN=1$ $SU(N)$ gauge theory coupled to $\CD_p[SU(N)]$ Argyres-Douglas theory with $N_f$ flavors and the certain superpotential $W=u_0$ shares the same fixed point as $\CN=1$ $SU(N)$ adjoint SQCD with $N_f$ flavors and $W_{A_{p}}=\Tr X^{p+1}$. We checked this duality by matching the chiral operators, central charges, and superconformal indices in a special limit. One open problem is that from the Lagrangian (or UV) sense, we cannot see how all of the chiral operators in the Argyres-Douglas side of the duality truncate, except for the special values of $p=2$ and $N=3$. While we believe that the truncation occurs more generally, it remains to be definitively shown. 


Our dual theory before turning on the superpotential $W=u_0$ has a geometric realization in $\CN=1$ class $\CS$ \cite{Gaiotto:2009we, Gaiotto:2009hg, Bah:2011vv, Bah:2012dg}. Namely, we can realize the $\CD_p[SU(N)]$ theory by considering the 6d $\CN=(2, 0)$ theory of type $A_{N-1}$ on a sphere with one maximal regular puncture and an irregular puncture of type $I_{N, N-p}$ \cite{Xie:2012hs}. Now, by gluing a three punctured sphere with two maximal and one minimal regular punctures along the maximal punctures with an $\CN=1$ vector multiplet, we obtain a three-punctured sphere with two (one maximal, one minimal) regular punctures and one irregular puncture (with the opposite color as the regular punctures \cite{Gadde:2013fma, Agarwal:2014rua}). This gives us the $\mathcal{T}_0$ theory with $N_f=N$. One can reduce the number of flavors $N_f<N$ by colliding two regular punctures to form an irregular puncture \cite{Bonelli:2011aa, Gaiotto:2012sf, Kanno:2013vi}. This construction hints towards a class $\CS$ description of the $W_{A_k}$ SCFTs.
Since the holographic duals of the  $\CD_p[SU(N)]$ SCFTs are now known \cite{Bah:2021mzw, Bah:2021hei}, one can also hope to utilize them to realize holographic duals of the $W_{A_{k}}$ SCFTs.

The $W_{A_k}$ SCFTs have a generalization labeled by Arnold's ADE simple surface singularities \cite{Intriligator:2003mi}. Therefore, it is natural to expect that there are generalizations of the $SU(N)$ gauge theory coupled to $\CD_p[SU(N)]$ sectors which flow to the $W_D$ and $W_E$ fixed points. Among them, we see hints that the $E_6$ and $E_8$ cases can be obtained by coupling two kinds of $\CD_p[SU(N)]$ theories, as follows. By considering $\CD_2[SU(N)]$ and $\CD_3[SU(N)]$ with $N$ being not divisible by $2$ and $3$, and gauging the diagonal $SU(N)$ while adding a number of fundamental flavors, we can trigger an RG flow to the $W_{E_6} = \Tr (Y^3 + X^4) $ fixed point upon adding a superpotential corresponding to the lowest dimension Coulomb branch operators of the $\CD_2[SU(N)]$ and $\CD_3[SU(N)]$ theories. Likewise, we can apparently obtain $W_{E_8} = \Tr(Y^3 + X^5)$ by considering $\CD_2[SU(N)]$ and $\CD_4[SU(N)]$ building blocks. It would be very interesting to explore these possible dualities further, especially as the $W_{E_6}$ and $W_{E_8}$ SCFTs do not have proposed Seiberg-like magnetic duals.  
On the other hand, the $W_{D_{k+2}} = \Tr (X^{k+1} + X Y^2)$ and $W_{E_7} = \Tr (Y^3 + Y X^3)$ cases do have proposed magnetic duals \cite{Brodie:1996vx,Kutasov:2014yqa}, but it is not currently clear how to obtain them from a deformation of Argyres-Douglas type theories. It would be interesting to find such realizations to further strengthen the connection between Argyres-Douglas theories and the $W_{ADE}$ fixed-point superconformal theories, as well as to explore possible subtleties with the $W_{D_{\text{even}}}$ and $W_{E_7}$ dualities, as per the discussion in \cite{Intriligator:2016sgx}. 

The adjoint SQCD fixed points provide several interesting insights on the nature of RG flow and strongly-coupled gauge theories. As we have discussed in this paper, they generically flow to product SCFTs with both an interacting sector and free sectors. This decoupling will modify the values of the central charges and indices on both sides of the dual theories. The decoupling happens only if the quantum effect is extremely strong, and it can significantly modify the IR physics, as in the examples studied in \cite{Barnes:2004jj, Maruyoshi:2018nod, Agarwal:2019crm, Agarwal:2020pol}. It would be interesting to further investigate the RG flows in this context with this point in mind, and especially to perform a refined check of unitarity beyond the chiral ring using the superconformal index \cite{Beem:2012yn, Evtikhiev:2017heo, Maruyoshi:2018nod}. 

We have also found new examples of supersymmetry enhancing RG flows between non-Lagrangian theories and Lagrangian gauge theories. It would be interesting to better understand such phenomenon, which may illuminate the vast landscape of non-Lagrangian field theories.

\begin{acknowledgments}
We thank Mykola Dedushenko, Monica Kang, Craig Lawrie and Ki-Hong Lee for helpful discussions. 
The work of KM is supported in part by JSPS Grant-in-Aid for Scientific Research No. 20K03935. The work of EN is supported in part by World Premier International Research Center Initiative (WPI), MEXT, Japan. The work of JS is supported in part by National Research Foundation of Korea Grant No. RS-2023-00208602 and also by POSCO Science Fellowship of POSCO TJ Park Foundation. JS and EN thank Seikei University for the hospitality where this work was conceived. 

\end{acknowledgments}

\appendix

\addtocontents{toc}{\protect\setcounter{tocdepth}{1}}

\section{SCFT conventions}
\label{sec:scft}

\subsection{$\CN=2$}

Let $r$ and $I_{a=1,2,3}$ denote the Cartan generators of the $U(1)_r\times SU(2)_R$ $\mathcal{N}=2$ R-symmetry.
An $\CN=2$ vector multiplet includes two Weyl fermions,  whose charges are given as $(r,I_3) = (1,\pm \tfrac{1}{2})$ with our chosen normalization, such that the 't Hooft anomalies are 
\ba{
\CN=2 \ \text{vector}:\qquad \tr \, r = \tr \, r^3 = 4 \tr \, r \, (I_3)^2 = 2 \, \text{dim}(G)\,.
}
An $\CN=2$ hypermultiplet in a representation $\CR$ includes two Weyl fermions with $(r,I_3) = (-1,0)$, such that its 't Hooft  anomalies are
\ba{
\CN=2 \ \text{hyper}:\qquad \tr \, r = \tr \, r^3 = - 2 \, \text{dim}( \CR)\,.
}
For an SCFT preserving $\CN=2$ supersymmetry, the anomaly coefficients are related to the $a$ and $c$ central charges as \cite{Kuzenko_2000}
\ba{
\tr\, r^3 = \tr\, r = 48 (a-c)\,,\qquad \tr \,r (I_3)^2 = 2 (2a-c)\,.
}

\subsection{$\CN=1$}

We can fix an $\CN=1$ subalgebra of the $\CN=2$ superconformal algebra, with  $U(1)_R$ symmetry generated by the combination, 
\ba{
\label{r0}
R_{0} = \frac{1}{3} r + \frac{4}{3} I_3\,.
}
With this choice, the following linear combination,
\ba{
\label{f0}
F = -r + 2 I_3\,,
} 
generates a $U(1)_F$ flavor symmetry from the $\mathcal{N}=1$ point of view.

An $\CN=1$ vector multiplet consists of one Weyl fermion with charges $(r,I_3) = (1,\tfrac{1}{2})$ in our chosen basis. The anomaly coefficients of the vector multiplet are thus
\ba{
\tr\, r = \tr \,r^3 =  \tr (2I_3)  = \tr (2I_3)^3 = \tr \,r (2 I_3)^2 = \tr \, r^2 (2 I_3) = \text{dim}(G)\,.
}

The $a$ and $c$ central charges are related to the 't Hooft anomalies for the superconformal  $U(1)_R$ R-symmetry as \cite{Anselmi_1998}
\ba{
a = \frac{3}{32} ( 3 \tr R_{\CN=1}^3 - \tr R_{\CN=1})\,,\qquad c = \frac{1}{32}( 9 \tr R_{\CN=1}^3 - 5 \tr R_{\CN=1})\,.
\label{acn1}
}
When the R-symmetry can mix with $U(1)$ flavor symmetries, the exact superconformal R-symmetry is given by locally maximizing the expression for $a$ in \eqref{acn1} over all possible $U(1)$ symmetries \cite{Intriligator:2003jj}.

\section{Facts about the \texorpdfstring{$\mathcal{D}_p[SU(N)]$}{DpSU(N)} SCFTs}
\label{sec:Dp}

In this appendix we review some of the features of the four-dimensional $\CN=2$ $\mathcal{D}_p[G]$ SCFTs. 

\paragraph{Class $\CS$ description}

The $\CD_p[G]$ theories have a Class $\CS$ description as the reduction of the 6d $(2,0)$ theory of type $G=ADE$ on a sphere with one full (maximal) puncture $F$, and one irregular puncture of type $G^{[h_G]}[p-h_G]$, for $h_G$ the dual Coxeter number of $G$ (using the notation of \cite{Xie:2012hs,Wang:2015mra}). We henceforth restrict to the case $G=A_{N-1}=SU(N)$, for which the irregular puncture is more commonly labeled by $A_{N-1}^{[N]}[k = p - N]$. The case $G=SU(2)$ is also known as $(A_1,D_p)$ in the class of $(G,G')$ theories which can be obtained from Type IIB string theory on a pair of hypersurface singularities \cite{ Cecotti:2010fi}; this class is furthermore  identified with the SCFT obtained by tuning to the maximal singular point on the Coulomb branch of pure $SO(2p)$ super Yang-Mills \cite{Argyres:2012fu}, as well as the maximal conformal point of $\CN=2$ $SU(p-1)$ gauge theory with 2 hypermultiplets. 

\paragraph{Operators}

As an $\CN=2$ SCFT, the $\CD_p[SU(N)]$ theories have an $SU(2)_R \times U(1)_r$ symmetry. 
  Their Coulomb branch is parameterized by  Coulomb branch operators that are neutral under $SU(2)_R$, and whose dimensions $\Delta = r/2$ are given by \cite{Xie:2012hs, Cecotti:2012jx, Cecotti:2013lda}
    \ba{
    \Delta
     =     \left[ j -  \frac{N}{p} s \right]_+ + 1\,,
           \label{UVdim}
    }
  where $\left[ x \right]_+ = x$ for $x>0$ and $0$ for $x\leq0$,
  and $j = 1,2,\ldots N-1$, $s=1,2,\ldots p-1$. The rank of the Coulomb branch is $\frac{1}{2}((N-1)(p-1)-\text{gcd}(p,N) + 1)$.
  For ${\rm gcd}(p, N)=1$, there are always Coulomb branch operators which we denote by $u_i$ with dimensions 
    \ba{
    \label{uidef}
    \Delta (u_i)
     =     \frac{p+1+i}{p}, ~~~i=0,1,\ldots p-2\,,
    }
  and other operators with higher dimensions. For instance, for $p=2$ and $N$ odd, there is one lowest Coulomb branch operator $u_0$ with dimension $3/2$, and a tower of higher operators with dimensions $\Delta = 3/2 + j$ for $j=1,\dots,(N-3)/2$. For $p=3$ and $\text{gcd}(3,N)=1$, there are two lowest Coulomb branch operators $u_0=4/3, u_1=5/3$, and then two towers with $\Delta = 4/3 + j_1$, $j_1=1,\dots, N/3\cdot (N \, \text{mod} \, 3) - 4/3$, and $\Delta = 5/3 + j_2$, $j_2=1,\dots, N/3\cdot (2N \, \text{mod}\,  3) - 5/3$.
  
  Each Coulomb branch multiplet --  whose primary Coulomb branch operator $u$ has $U(1)_r$ charge $r$ and dimension $\Delta(u) = r/2$ -- contains a level-two descendant scalar operator $v$ with $U(1)_r$ charge $r-2$, $SU(2)_R$ charge $I_3=1$, and dimension $\Delta(v) = r/2 + 1$.  We will denote these descendants of the  $u_{i}$ defined above in \eqref{uidef} as $v_{i}$.

  The flavor symmetry of the SCFT is at least $SU(N)$, with $\text{gcd}(p,N)-1$ extra $U(1)$'s for ${\rm gcd}(p, N) \neq1$. 
  We will focus on the case with ${\rm gcd}(p, N)=1$. Associated with the flavor symmetry, there exists a conserved current multiplet whose lowest component is the moment map operator $\mu$ with $r=0$ and $I_3=1$, and $\Delta(\mu)=2$. 
  It was argued in \cite{Agarwal:2018zqi} that this operator satisfies the following chiral ring relations,
    \ba{
    \tr \mu^k  =     0\,,\qquad
    \mu^p \big|_{{\rm adj}}=     0\,,
    \label{chiralringDpG}
    }
  for any $k$, where $\ldots|_{{\rm adj}}$ denotes the adjoint part of $\ldots$.

\paragraph{SCFT data}

  The $SU(N)$ flavor central charge is given by
    \ba{
    k_{SU(N)}
     =     \frac{2(p-1)}{p} N\,,
    }
  defined by $-2 {\rm Tr}\, r T^a T^b = k_{SU(N)} \delta^{ab}$,
  where $T^a$ are the generators of the $SU(N)$ global symmetry.
  The $a$ and $c$ central charges of the $\mathcal{D}_p[SU(N)]$ theory for coprime $p,N$ are given by
    \ba{
    a  =     \frac{(4p-1)(p-1)}{48p} (N^2-1)\,, \qquad
    c  =     \frac{p-1}{12} (N^2-1)\,.
    \label{acdpg}
    }
    The non-zero 't Hooft anomaly coefficients for the R-symmetry are
    \ba{
    \tr\, r =\tr\, r^3 = \frac{(1-p)(N^2-1)}{p}\,,\qquad \tr\, r I_3^2 = \frac{(1-p)(1-2p)(N^2-1) }{12p}\,.
    }

\bibliographystyle{jhep}
\bibliography{MNSbib}

\end{document}